\documentclass[
 reprint,
 nofootinbib,
 preprintnumbers,
 amsmath,amssymb,
 aps,
 prd
]{revtex4-2}

\usepackage{dcolumn}% Align table columns on decimal point
\usepackage{bm}
\usepackage{appendix}
\usepackage{amsmath}
\usepackage{slashed}
\usepackage{multirow}
\usepackage{color}
\usepackage{mathtools}
\usepackage{graphicx}
\usepackage{braket}
\usepackage{setspace}
\usepackage{graphicx}
\usepackage{geometry}
\usepackage{hyperref}
\usepackage{starfont}
\usepackage{amssymb}
\usepackage{bbold}
\usepackage[shortlabels]{enumitem}
\usepackage{framed}
\usepackage{empheq}
\usepackage{amsmath}
\usepackage{tikz-feynman}
\usepackage{simpler-wick}
\usepackage{wasysym}
\usepackage{mhchem}
\usepackage{extarrows}
\usepackage{aas_macros}

\DeclarePairedDelimiter\abs{\lvert}{\rvert}
\makeatletter
\let\oldabs\abs
\def\abs{\@ifstar{\oldabs}{\oldabs*}}
\makeatother

\def\bea{\begin{eqnarray}}
\def\eea{\end{eqnarray}}

\newcommand{\vev}[1]{\left\langle #1\right\rangle}

\def\eV{\,{\rm eV}}

\def\tr{\,{\rm tr}}

\def\bfx{{\bf x}}

\def\bfy{{\bf y}}

\def\bfk{{\bf k}}
\def\bfp{{\bf p}}

\def\bfv{{\bf v}}

\def\sch{{Schr{\"o}dinger }}

\begin{document}
\preprint{DESY-21-217}

\title{Gravitational Focusing of Wave Dark Matter}
\author{Hyungjin Kim}
\email{hyungjin.kim@desy.de}
\affiliation{%
Deutsches Elektronen-Synchrotron DESY, Notkestr.\,85, 22607 Hamburg, Germany}
\author{Alessandro Lenoci}
\email{alessandro.lenoci@desy.de}
\affiliation{%
Deutsches Elektronen-Synchrotron DESY, Notkestr.\,85, 22607 Hamburg, Germany}

\begin{abstract}
A massive astrophysical object deforms the local distribution of dark matter, resulting in a local overdensity of dark matter. 
This phenomenon is often referred to as gravitational focusing.
In the solar system, the gravitational focusing due to the Sun induces modulations of dark matter signals on terrestrial experiments.
We consider the gravitational focusing of light bosonic dark matter with a mass of less than about 10 eV. 
The wave nature of such dark matter candidates leads to unique signatures in the local overdensity and in the spectrum, both of which can be experimentally relevant. 
We provide a formalism that captures both the gravitational focusing and the stochasticity of wave dark matter, paying particular attention to the similarity and difference to particle dark matter. 
Distinctive patterns in the density contrast and spectrum are observed when the de Broglie wavelength of dark matter becomes comparable or less than the size of the system and/or when the velocity dispersion of dark matter is sufficiently small. 
While gravitational focusing effects generally remain at a few percent level for a relaxed halo dark matter component, they could be much larger for dark matter substructures.
With a few well-motivated dark matter substructures, we investigate how each substructure responds to the gravitational potential of the Sun.
The limit at which wave dark matter behaves similar to particle dark matter is also discussed.
\end{abstract}

\maketitle
%\tableofcontents
%change the config file and put the preprint command

\section{Introduction}

The evidence for dark matter (DM) has been accumulating for decades.
Since the evidence is based on the gravitational interaction, it still remains a mystery how dark matter communicates with the standard model. 
While there exists a plethora of possibilities, a particular class of DM candidates, namely weakly interacting massive particle, has given particular attention due to its predictive power and theoretical motivations.
However, it has avoided intense scrutiny of terrestrial DM searches as well as cosmological and astrophysical probes.  

A light bosonic particle with a mass smaller than $10\eV$ provides a compelling alternative dark matter candidate.
Such a DM candidate arises from various beyond the standard models with different theoretical and phenomenological motivations. 
One specific example is the QCD axion~\cite{Preskill:1982cy, Abbott:1982af, Dine:1982ah}, which is originally proposed as a solution to the strong CP problem in the standard model~\cite{Peccei:1977hh, Weinberg:1977ma, Wilczek:1977pj, Kim:1979if, Shifman:1979if,  Dine:1981rt, Zhitnitsky:1980tq}. 
Not only from a solution to the strong CP problem, it also arises from dynamical and anthropic solutions to the electroweak hierarchy problem~\cite{Graham:2015cka, Arvanitaki:2016xds, Banerjee:2018xmn, Arkani-Hamed:2020yna, TitoDAgnolo:2021nhd, TitoDAgnolo:2021pjo}.
Moreover, a large number of light bosonic states may arise from higher form fields on compact extra dimensions in some ultraviolet theories~\cite{Svrcek:2006yi, Arvanitaki:2009fg}, and one of them can be  dark matter in the present universe. 

A light bosonic DM candidate behaves like a classical wave rather than an individual particle. 
With a typical value of DM mass density in a collapsed halo, the number of DM particles within a volume of de Broglie wavelength takes a gigantic value.
This allows us to treat DM as a collection of waves, and hence, we call such candidates wave dark matter candidates.
The wave nature makes them qualitatively different from usual particle DM candidates, leaving unique cosmological and astrophysical signatures throughout the entire history of the universe. 
See a recent review~\cite{Hui:2021tkt} for an overview.

As it appears ubiquitously in many beyond the standard models, many efforts have been put forward to find such wave dark matter in Milky Way for past decades (see e.g.~\cite{Graham:2015ouw} for a review on axion and axion-like particle searches). 
When it comes to terrestrial searches of dark matter, one essential input is the dark matter local distribution.
A common starting point is to assume that the dark matter forms an isothermal halo and its velocity distribution is given by the Maxwell-Boltzmann distribution~\cite{Drukier:1986tm, Freese:1987wu}.
Such a model is referred to as the standard halo model, and as its name implies, it has been the standard assumption for various analyses of terrestrial dark matter searches, not only for the wave dark matter but also for particle dark matter. 
While the standard halo model is simple and is reproduced reasonably well from N-body studies~\cite{Vogelsberger:2008qb, Kuhlen:2009vh}, it ignores several important aspects of our current understanding of the local dark matter distribution, which might be important to determine a precise map of local dark matter. 

Firstly, it does not account for the potential existence of DM substructures near the solar neighborhood. 
With precision astrometric data, recent analyses have identified a stellar population near the solar neighborhood with a distinct kinematic structure~\cite{2018MNRAS.478..611B, 2018Natur.563...85H}.
This stellar substructure is called Gaia-Enceladus or Gaia-Sausage due to its elongated structure in the velocity space. 
It is inferred that this stellar substructure was created through a recent merger with a satellite galaxy of mass $\sim 10^{10}M_\odot$ about $8\textrm{--}10$~Gyr ago.
The same merger event is expected to deposit DM into the Milky Way.
Numerical studies showed that accreted dark matter from the same merger event has a kinematic structure very similar to its stellar counterpart~\cite{Necib:2018iwb, Necib:2018igl}.
A nontrivial fraction of dark matter in the inner halo is therefore expected to be in form of substructures~\cite{Evans:2018bqy, OHare:2018trr, OHare:2019qxc}. 
In addition to Gaia-Enceladus, there could be other substructures with completely different kinematic properties.
Examples include streams with characteristically small velocity dispersion, and a dark disk which may arise from the merger events that are responsible for the formation of the thick stellar disk in the Milky Way~\cite{2008MNRAS.389.1041R}. 

Secondly, it does not account for the deformation of local dark matter distribution due to the gravitational potential of the Sun. 
As we are bound to the solar system, all the obervations are eventually affected by the presence of the Sun. 
Specifically, the gravitational potential of the Sun deflects the trajectory of DM, gravitationally focusing DM particles in the solar system.
This leaves a tail of local overdensity in the opposite direction to the direction of the solar system in the Milky Way. 
It is therefore important to examine how the dark matter profile near the solar system is modified by the Sun, especially along the orbital trajectory of the Earth.
This gravitational focusing of DM has been studied, especially in the context of weakly interacting massive particle DM searches~\cite{1957MNRAS.117...50D, 1967AJ.....72..219D, Griest:1987vc, Sikivie:2002bj, PhysRevD.74.083518, Lee:2013wza}. 
While it was found that the event rate due to the local overdensity remains only a few percent level in the standard halo model, it can lead to an interesting pattern of DM flux in the sky map~\cite{Sikivie:2002bj, PhysRevD.74.083518} as well as modifying the DM annual modulation signal~\cite{Lee:2013wza}. 

The purpose of this work is to extend the discussion to the gravitational focusing of wave DM and to examine how wave dark matter substructures are focused in the solar system.
We consider a spinless, light, bosonic dark matter candidate, minimally coupled to the gravity, with a mass smaller than $10\,\eV$. 
We first develop a formalism which allows us to study the gravitational focusing and the stochasticity of wave dark matter simultaneously. 
Unlike particle DM, systems of wave DM include an additional spatial scale characterized by de Broglie wavelength, and much of wave properties of the problem is controlled by the de Broglie wavelength; for instance, the overdensity becomes constant when the radial distance from an astrophysical source becomes smaller than de Broglie wavelength and the width of density wake field is controlled by de Broglie wavelength.
All of these features are distinct from particle dark matter and they are commonly observed in the other context of wave dark matter studies at different scales; for instance the same wave nature flattens the inner profile of DM halo, and hence, may resolve small-scale issues in the cold dark matter paradigm~\cite{Hu:2000ke}. In addition, this wave nature reduces the dynamical friction, friction induced by the density wake field on astrophysical objects~\cite{Hui:2016ltb, Bar-Or:2018pxz, Lancaster:2019mde}. 
We discuss below how this wave nature manifests itself in the solar system and also discuss possible implications on terrestrial experiments.

The paper is organized as follows. 
In Section~\ref{sec:focusing_particle}, we briefly review the gravitational focusing of particle DM for the purpose of later comparison with the wave DM. 
We compute the local density contrast for the particle DM and discuss its behavior as a function of parameters of the system, such as DM mean velocity and dispersion. 
In Section~\ref{sec:focusing_wave}, we investigate the gravitational response of wave DM. 
The investigation is done in several steps. 
We first begin with an action of a minimally coupled spinless bosonic particle, and expand the field in terms of creation and annihilation operator with a mode function. 
Then, we solve the \sch equation for the mode function in Section~\ref{sec:wave_function}, which encodes an information on the gravitational potential and spatial distribution of the wave DM. 
We then determine statistical properties of the DM field by specifying the density operator of the system, which is discussed in  Section~\ref{sec:stat}. 
Specification of density operator allows us to compute the ensemble average of arbitrary operators constructed from the field. 
This procedure together with the mode function obtained from the \sch equation enables us to consider the gravitational focusing and the stochasticity of the field at the same time. 
After discussing ensemble averages of the field, we compare the wave gravitational focusing with particle dark matter focusing, contrasting the similarities and differences between them. 
In Section~\ref{sec:application}, we apply the formalism discussed in the previous section to the DM substructures in the solar system as well as relaxed halo dark matter to see how each substructure responds to the gravitational potential of the Sun. 
We discuss the effect of coordinate transformation as well as the classical limit of wave DM in Section~\ref{sec:discussion} and conclude in Section~\ref{sec:conclusion}. 
Throughout the work, we use natural unit $c=\hbar=1$ except for the section we discuss the classical limit of wave DM.

%% Particle GF %%
\section{Particle focusing}\label{sec:focusing_particle}
We review the gravitational focusing of particle dark matter.
We follow~\cite{Sikivie:2002bj, PhysRevD.74.083518}. 
The results summarized in this section are to compare the particle DM result with that of wave DM. 
Readers who are already familiar with the gravitational focusing of particle DM may continue to Section~\ref{sec:focusing_wave}. 

\begin{figure}[t]
\includegraphics[width=0.4\textwidth]{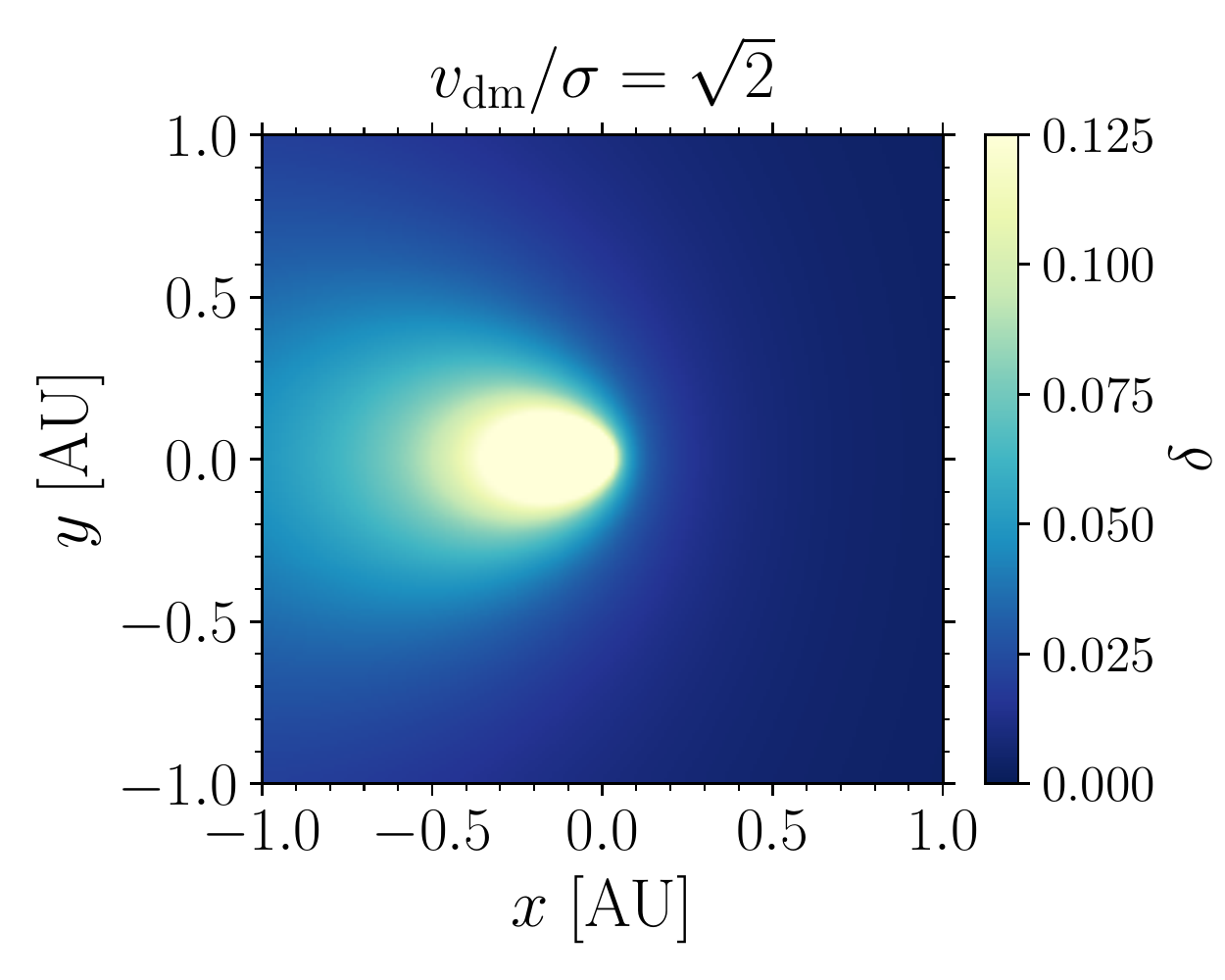}
\caption{We show the density contrast of the particle dark matter.
We choose $\bfv_{\rm dm}=(-240,0,0)$~km/s, $v_{\rm dm}/ \sigma =\sqrt{2}$ and the source mass to be the solar mass.}
\label{fig:delta_classic}
\end{figure}   

Let us consider the dark matter phase space distribution, $f(t,\bfx,\bfp)$.
In the collisionless system, the phase space density is conserved, i.e. $df/dt =0$.
According to Liouville's theorem, the phase space volume should be conserved along the trajectory given by the Hamiltonian of the system.

The dark matter local overdensity can be easily estimated by Liouville's theorem. 
Consider particles confined in an infinitesimal phase space volume, $d^3 v_0 d^3 x_0$, asymptotically far away from a source of mass $M$. 
For simplicity, let us assume a constant phase space density over this infinitesimal volume.
Due to the gravitational potential, particles are accelerated and their speed at a radial distance $r$ becomes $v^2(r) = v_0^2 + 2 GM /r$, where $v_0$ is the initial speed far away from the source $M$. 
According to Liouville's theorem, the phase space volume remains constant, $d^3 x_0 \, d^3 v_0 = d^3 x \, d^3 v$, where $d^3x \, d^3v$ is the phase space volume of particles at $r$.
This leads to $d^3x = [v_0 / v(r)] d^3x_0$.
We find that the particle mass density at $r$ is 
\bea
\rho(r) \approx \rho_0 [1 + v_e^2(r) / v_0^2]^{1/2} , 
\label{est_density_contrast}
\eea
where $v_e^2(r) = 2GM/r$ is the escape velocity and $\rho_0$ is the mass density far away from the source $M$. 

More quantitative analysis is possible. 
Suppose that $f(\bfv_0)$ is the phase space distribution at $r\to\infty$, normalized as $\int d^3 v_0 \, f(\bfv_0) =1 $. 
The differential particle mass in an infinitesimal phase space volume is $m dN=  \rho_0 f(\bfv_0) d^3x_0 d^3 v_0  =  \rho_0  f(\bfv_0) d^3 x d^3 v$.
The mass density at the position $\bfx$ relative to its asymptotic value is given by
\bea
\frac{\rho(\bfx)}{\rho_0} 
= 1 + \delta
&=& \int d^3 v\, f(\bfv_0)
\nonumber\\
&=& \int d^3 v_0\, d(\bfv_0,\bfx) f(\bfv_0).
\label{particle_overdensity}
\eea
where $\delta$ is the density contrast. 
The Jacobian $d(\bfv_0,\bfx)$ is~\cite{Sikivie:2002bj}
$$
d(\bfv_0,\bfx) = 
\frac{1}{2} 
\left[ 
\left(1 + \frac{2v_e^2 /v_0^2}{ 1 -  \hat v_0 \cdot \hat x } \right)^{\frac12}
\!\! + 
\left(1 + \frac{2v_e^2 /v_0^2}{ 1 -  \hat v_0 \cdot \hat x } \right)^{-\frac12}
\right].
$$
For a given velocity distribution, the above integral can be numerically performed. 
One can directly compute the first integral by using an explicit expression of $\bfv_0 = \bfv_0(\bfv,\bfx)$~\cite{PhysRevD.74.083518}. 

For the further discussion, we consider the Maxwell-Boltzmann distribution function $f(\bfv)$,
\bea
f(\bfv) = \frac{1}{(2\pi\sigma^2)^{3/2}} \exp\left[ - \frac{(\bfv - \bfv_{\rm dm})^2}{2\sigma^2} \right].
\label{mb}
\eea
The velocity distribution is characterized by the mean velocity is $\langle \bfv \rangle = \bfv_{\rm dm}$ and the velocity dispersion  $\sigma$. 
With the Maxwell-Boltzmann distribution, the density contrast is completely described  by three parameters,
$$
v_{\rm dm}/\sigma, \qquad 
r/\bar{r}, \qquad 
\hat{x}\cdot \hat{v}_{\rm dm} (\equiv \mu ),
$$
where  
$$
\bar{r} = \frac{GM}{\sigma^2}= 0.03\; {\rm AU}\;\left(\frac{M}{M_\odot}\right)\left(\frac{240 \;{\rm km/sec}}{\sqrt{2}\sigma}\right)^2
$$
is the radius of gravitational influence. 
One may choose $v_e(r)/\sigma$ instead of $r/\bar{r}$. 
We present the density contrast of particle dark matter in Figure~\ref{fig:delta_classic} with $\bfv_{\rm dm} = (-240,\, 0 , \, 0)$~km/sec and $v_{\rm dm}/ \sigma = \sqrt{2}$ for the purpose of demonstration.
The density contrast is invariant under the rotation with respect to $\hat{v}_{\rm dm}$-axis due to an azimuthal symmetry.

A characteristic density wake field is formed along $\hat{v}_{\rm dm}$. 
Two variables might be useful to inspect the density wake field,
\bea
\delta_{\rm ds} = \delta|_{\mu =1},
\qquad
\delta_{\rm avg} = \int \frac{d\Omega_{\hat x}}{4\pi} \delta
= \int \frac{d\mu}{2} \delta,
\eea
where $\delta_{\rm ds}$ is the density contrast along the downstream ($\hat{x}\cdot\hat{v}_{\rm dm} =1$) and $\delta_{\rm avg}$ is the density contrast averaged over a sphere of radius $r$.  
While above variables are a function of two parameters, $v_{\rm dm}/\sigma$ and $r/\bar{r}$, they practically depend only on one specific combination of these two.
For the downstream density contrast, we find
\bea
1 + \delta_{\rm ds} 
\approx \Big( 1  + \frac{v_e^2}{\sigma^2} \Big)^{\frac12}
=  \Big( 1  + \frac{2 \bar{r}}{r} \Big)^{\frac12}.
\label{particle_down_est}
\eea
while, for the averaged density contrast, we find
\bea
1 + \delta_{\rm avg} 
&=& \int d^3 v_0\, f(\bfv_0) \big(1+ \frac{v_e^2}{v_0^2}\big)^{\frac12}
\nonumber\\
&\approx&
\Big( 1 + \frac{v_e^2}{\sigma^2+ v_{\rm dm}^2} \Big)^{\frac12} .
\label{particle_avg_est}
\eea
We check these expressions numerically in Appendix~\ref{app:particle}. 
The numerical results are within a factor of two for wide range of parameters.
Note also that the ratio $\delta_{\rm avg}/\delta_{\rm ds}$ represents the angular scale over which the density contrast takes nonvanishing value. 
For $\sigma > v_e$ and $v_{\rm dm} > \sigma$, this angular scale is approximately given by $\Delta \mu \sim (\sigma / v_{\rm dm})^2$.

%% Wave focusing %%
\section{Wave focusing}\label{sec:focusing_wave}
We consider the wave dark matter in this section.
For concreteness, we consider a spinless bosonic dark matter with a mass $m$, minimally coupled to the gravity. 
We approach the problem in several steps.
We begin with the action and expand the field in terms of creation and annihilation operators with a mode function. 
A mode function for the system is obtained by solving the equation of motion, which is discussed in Section~\ref{sec:wave_function}. 
The mode function obtained in this way contains information about spatial distribution of the wave near the source $M$. 
Statistical properties of the field are determined by specifying the underlying density operator of the system, which is discussed in Section~\ref{sec:stat}.

The action of the system is 
\bea
S = \int d^4x \, \sqrt{-g} 
\left[
\frac{1}{2} g^{\mu\nu} \partial_\mu \phi \partial_\nu \phi - \frac{1}{2} m^2 \phi^2
\right], 
\eea
where $g = \det g_{\mu\nu}$ is the determinant of the metric, $g_{\mu\nu}$ is the metric tensor, and $m$ is the dark matter mass. 
The metric is given as 
\bea
ds^2 =( 1 + 2 \Phi) dt^2 - (1-2\Phi) d\vec x\,^2.
\eea
where $d\vec x\,^2=dr^2+r^2d\Omega$ and $\Phi$ is the gravitational potential. Since we are interested in non-relativistic dark matter, we expand the field as
\bea
\hat{\phi}(x) =  \sum_i \frac{1}{\sqrt{2 m V}}  
\Big[ 
	\hat{a}_i \psi_i (x) e^{-i m t} +
	\hat{a}_i^\dagger \psi_i^* (x) e^{i m t} 
\Big] ,
\label{phi_ex_nonrel}
\eea
where $\hat a_i$ and $\hat a_i^\dagger$ are an annihilation and creation operator, satisfying the canonical commutation relation $[\hat a_i, \hat a_j^\dagger] = \delta_{ij}$. 
We have expanded the field in a finite box of volume $V$.
The index $i$ denotes all possible quantum numbers.
We solve the equation of motion to determine the mode function $\psi_i(x)$ and specify the density operator in the following sections as advertised.

%% WAVE FUNCTION %%
\subsection{Wave function}\label{sec:wave_function}
We first solve the equation of motion for the mode function. 
The wave function is given by the solution to the Klein-Gordon equation, $(\square + m^2 ) \phi = 0$. 
In the nonrelativistic liimt, we expand the field as in Eq.~\eqref{phi_ex_nonrel}. 
In the limit $|\ddot{\psi}/\dot{\psi}|, \, |\dot\psi / \psi| \ll m$ and $|\dot\Phi/\Phi| \ll m$, the Klein-Gordon equation is reduced to the \sch equation,
\bea
i \partial_t \psi = \Big( - \frac{\nabla^2}{2m} + m \Phi \Big) \psi.
\label{sch}
\eea
The gravitational potential $\Phi$ is determined by the Poisson equation, $\nabla^2 \Phi = 4\pi G (\rho_M + \rho_{\rm dm})$, where $\rho_M$ and $\rho_{\rm dm}$ are the mass density of the source and dark matter, respectively. 
Ignoring the self-gravity, the gravitational potential is approximated as $\Phi(r) = - GM /r$. 
In this case, solving the \sch equation Eq.~\eqref{sch} is identical to solving the Coulomb scattering problem of an electron under the central $1/r$ potential generated by the proton. 
With $\psi_i (x) \to e^{-i (k_i^2/2m) t} \psi_i(\bfx)$, the wave function is given as~\cite{Landau:1991wop, Hui:2016ltb}
\bea
\!\!\!\!\!\!\!\!
\psi_{\bfk}(\bfx) \! = \!
e^{i \bfk\cdot\bfx} \Gamma(1 -i \beta)e^{\pi\beta/2}
M\big[i\beta,1,ikr(1-\hat{k}\cdot\hat{x})\big]. 
\eea
where $\beta= G Mm^2  /k $ and $M(a,b,c)$ is the confluent hypergeometric function.
Each mode is characterized by the wavenumber $\bfk$.
For $k r \gg 1$ and $\mu = \hat k \cdot \hat x < 0$, the wave function is approximated a plane wave $\psi_{\bfk}(\bfx) \sim e^{i\bfk \cdot \bfx}$.

%% STAT %%
\subsection{Statistical properties}\label{sec:stat}
The discussion in the previous section shows how a plane wave of wavenumber $\bfk$ responds to the gravitational potential of the source $M$. 
A realistic dark matter halo is a superposition of different modes, and so is the field $\hat{\phi}$. 
To determine the statistical properties of $\hat{\phi}$, we must specify the density operator of the system.

\subsubsection{Density operator}
Before we discuss the density operator for the field, let us begin with a simple harmonic oscillator in quantum mechanics as a toy model.
The statistical properties of simple harmonic oscillator are described by the density operator $\hat{\rho}$. 
If the system is in the thermal equilibrium, the density operator is given by $\hat\rho = e^{-\beta \hat{H}} / \tr ( e^{-\beta \hat H})$ with a Hamiltonian of the system $\hat{H} = \omega a^\dagger a$.
Alternatively, the density matrix can be written as
\bea
\hat\rho = \frac{1}{1+\langle n \rangle} \sum_n  \left( \frac{\langle n \rangle}{\langle n \rangle + 1 } \right)^n |n  \rangle \langle n |,
\label{den_qm}
\eea
where $|n\rangle$ is an eigenstate of the number operator $\hat N = a^\dagger a$. 
The mean occupation number is $\langle n \rangle = (e^{\beta \omega} -1)^{-1}$. 
While the above form is obtained for the thermal system, it applies for much broader systems as long as the statistical properties of the excitation are chaotic~\cite{1973qtl..book.....L}.
More specifically, the above density matrix maximizes the entropy of the system for a fixed mean occupation number $\langle n \rangle$.
Without assuming thermal equilibrium, we use the above form of density matrix for the following discussion. 

Since we are interested in a system with a large occupation number, it is convenient to express the above density matrix in the coherent state representation. 
The coherent state is defined as $| \alpha \rangle = e^{-|\alpha|^2/2} \sum \frac{\alpha^n}{ \sqrt{n!} } | n \rangle$ with a complex number $\alpha$.
It is an eigenstate of the annihilation operator, $\hat a |\alpha \rangle = \alpha | \alpha \rangle$. 
With the coherent state, the density operator can be expressed~\cite{PhysRev.131.2766, PhysRevLett.10.277},
\bea
\hat \rho &=& 
\int d^2\alpha \,
P(\alpha) | \alpha \rangle \langle \alpha |.
\label{single_mode_density_coherent}
\eea
where $P(\alpha)$ is a quasi-probability distribution, defined as
\bea
P(\alpha) = \frac{1}{\pi \langle n \rangle} 
\exp\left[ - \frac{|\alpha|^2}{\langle n \rangle } \right]. 
\label{prob}
\eea
For the density matrix Eq.~\eqref{den_qm}, the quasi-probability distribution has desired properties to be interpreted as a probability distribution for a complex number $\alpha$; it is positive, $P(\alpha) \geq 0$, and it is integrated up to unity, $\int d^2\alpha \,  P(\alpha) =1 $. 
Moreover, the quasi-probability distribution can be factorized into two probability distributions, one for the modulus $|\alpha|$ and one for the phase $\theta = {\rm arg}(\alpha)$,
$$
P(\alpha) = P(|\alpha|) P(\theta)
$$
where the modulus $|\alpha|$ follows the Rayleigh distribution, $P(|\alpha|) = 2 ( | \alpha | / \langle n \rangle) \exp[ - |\alpha|^2 / \langle n \rangle ] $, while the phase $\theta$ follows the uniform distribution, $P(\theta) = 1/2\pi$. 
The expectation value of any operator $\hat {\cal O}(\hat a, \hat a^\dagger)$ is given by $\langle \hat{\cal O} \rangle = \tr (\hat{\cal O} \hat \rho) = \int (d^2 \alpha / \pi) \langle \alpha | \hat{\cal O} \hat{\rho} | \alpha \rangle$. 
Since the coherent state is the eigenstate of annihilation operator, we may replace $\hat{\cal O}(\hat a , \hat a^\dagger) \to {\cal O}(\alpha , \alpha^*)$ and compute the expectation value with the probability distribution $P(\alpha)$,
\bea
\langle \hat{\cal O}(\hat a, \hat a^\dagger) \rangle
\approx 
\int d^2\alpha \,P(\alpha) {\cal O}(\alpha, \alpha^*) . 
\label{expectation_qm}
\eea
The only subtlety is that the commutation of ladder operators does not vanish, leading to additional terms in the above expression. 
However, those additional contributions are suppressed by powers of the mean occupation number, and therefore, they are subleading as long as the occupation number is large.
The error of the above approximation is ${\cal O}(1/\langle n \rangle)$.

Motivated by this, we take the same form of density matrix for the system of  the DM field $\hat \phi$. 
Since we decomposed the field as a linear combination of different modes, the density operator is now given as product of Eq.~\eqref{single_mode_density_coherent} for each mode $i$,
\bea
\hat\rho 
= 
\bigg[
\prod_i \int d^2 \alpha_i \, P(\alpha_i) 
\bigg]
| \{ \alpha_i \} \rangle \langle \{ \alpha_i \} |,
\eea
where $P(\alpha_i)$ is the probability distribution for each $\alpha_i$.
The form of $P(\alpha_i)$ is the same as in Eq.~\eqref{prob}. 
The coherent state here is defined as $\hat a_j | \{\alpha_i\} \rangle = \alpha_j | \{\alpha_i \} \rangle.$

The density operator completely determines the statistical properties of the scalar field. 
We can compute the expectation value of any operators constructed from $\hat\phi$.
Similarly to the quantum mechanical harmonic oscillator, one may replace $\hat{a}_i \to \alpha_i$, or $\hat{\cal O}(\hat{a}_i, \hat{a}_i^\dagger) \to \hat{\cal O}(\alpha_i, \alpha^*_i)$, and treat each $\alpha_i$ as a random variable following the probability distribution $P(\alpha_i)$. 
As in the quantum mechanics example, there are additional terms arising from nonvanishing commutation relation between creation and annihilation operators, but such contributions are again suppressed by powers of the mean occupation number.
For instance, we find 
\bea
\langle a_i^\dagger a_j \rangle &=& \delta_{ij} \langle n_i \rangle,
\\
\langle a_j a_i^\dagger  \rangle &=& \delta_{ij} ( \langle n_i \rangle +1)
\approx \delta_{ij} \langle n_i \rangle,
\eea
where we see the quantum fluctuation is suppressed by $1/\langle n_i\rangle$. 
This allows us to write the quantum field as
$$
\hat \phi(t,\bfx) 
\to \phi(t,\bfx) 
= 
\sum_i \frac{1}{\sqrt{2 m V}}  
\Big[ 
	\alpha_i \psi_i (\bfx) e^{-i \omega_i t} + {\rm h.c.}
\Big] ,
$$
where $\omega_ i = m + k_i^2/ 2m$. 
We treat each $\alpha_i$ as a random variable, following the probability distribution Eq.~\eqref{prob}, while ignoring possible quantum fluctuation suppressed by $1/\langle n \rangle$.
In the plane wave limit, $\psi_{\bfk}(\bfx) = e^{i\bfk\cdot\bfx}$, this approach easily reproduces the statistical properties of the scalar field $\hat{\phi}$ obtained in different approaches~\cite{Derevianko:2016vpm, Foster:2017hbq}.
Including these stochastic properties is important for data analysis of wave DM experiments as it may affect interpretation of data~\cite{Derevianko:2016vpm, Foster:2017hbq, Centers:2019dyn, Lisanti:2021vij, Gramolin:2021mqv}.

% MOMENTs AND SPECTRUM %
\subsubsection{Moments and spectrum}
Equipped with the density matrix and the wave function, we proceed to compute the ensemble averages, i.e. moments, of the field $\hat{\phi}$.
We are particularly interested in the second moment of the field for two reasons.
The second moment of the field $\langle \hat\phi^2 \rangle$ is directly related to the density contrast as $1 + \delta = \langle \hat\phi^2 \rangle / \phi_0^2$, where $\phi_0$ is the root mean square of the field value measured at asymptotically far away from the source. 
From this, we can easily compare the density contrast of wave and particle dark matter.
Secondly, this quantity is, in many cases, directly related to the observables in the terrestrial detectors. 
For instance, in axion and axion-like particle searches for ${\cal L} \supset \phi F \widetilde F$,\footnote{For instance, ADMX~\cite{ADMX:2018gho, ADMX:2019uok, ADMX:2021nhd}, HAYSTAC~\cite{Brubaker:2016ktl, HAYSTAC:2018rwy}, ABRACADABRA~\cite{Kahn:2016aff, Ouellet:2018beu}, DM-Radio~\cite{Chaudhuri:2018rqn}, CAPP~\cite{CAPP:2020utb}, MADMAX~\cite{Caldwell:2016dcw}, ORGAN~\cite{McAllister:2017lkb} and so on.} a total signal power is often proportional to $\langle \hat \phi^2 \rangle$ and signal power spectral density is related to the spectral density of $\langle \hat \phi^2 \rangle$.
The same holds for scalar dark matter searches on ${\cal L} \supset \phi \bar \psi \psi,\, \phi FF$ via the oscillations of fundamental constants.\footnote{For instance, atomic clocks~\cite{Arvanitaki:2014faa, VanTilburg:2015oza, Hees:2016gop, Aharony:2019iad, Kennedy:2020bac, Savalle:2020vgz, Campbell:2020fvq}, atomic and molecular spectroscopy searches~\cite{Antypas:2019qji, Oswald:2021vtc}, and interferometry~\cite{Arvanitaki:2016fyj, Grote:2019uvn, Vermeulen:2021epa, Aiello:2021wlp, Badurina:2021lwr}.}

Using a generalized expression of Eq.~\eqref{expectation_qm}, we obtain the variance of the field as
\bea 
\vev{\hat \phi^2(t,\bfx)} 
&=&  \int\frac{d^3 k}{(2\pi)^3}\frac{n_\bfk}{m}|\psi_\bfk(\bfx)|^2 
\nonumber \\
&=& \phi_0^2 \int d^3 v\; f(\bfv) |\psi_\bfk(\bfx)|^2\ ,
\eea
where we have taken the continuum limit in the first line.
Here, $\phi_0$ is the mean field value at asymptotically far away from the source and $n_{\bfk}$ is the occupation number, which is the continuum version of $n_i$ defined in the density matrix. 
We have also defined the velocity probability distribution function $f(\bfv) = (m/2\pi)^3 n(\bfk) / (m \phi_0^2)$.
The asymptotic field value $\phi_0$ is related to the asymptotic mass density as $\phi_0 = \sqrt{\rho_0}/m$.
The variance of the field is directly related to the density contrast as
\bea
1 + \delta = \frac{\langle \phi^2 \rangle}{\phi_0^2} 
=\int d^3 v\,  f(\bfv) |\psi_\bfv(\bfx)|^2 .
\label{wave_delta}
\eea
This expression can be compared with the corresponding expression for particle dark matter, Eq.~\eqref{particle_overdensity}. 
The Jacobian factor in Eq.~\eqref{particle_overdensity} that describes the particle flow is now replaced with the squared amplitude of the wave function.
For the Maxwell-Boltzmann distribution Eq.~\eqref{mb}, the density contrast for the wave dark matter is controlled by four parameters
\bea
v_{\rm dm} / \sigma, \qquad
r/\bar{r}, \qquad
\hat{x} \cdot \hat{v}_{\rm dm}, \qquad
 2\pi \bar{r} / \lambda_{\rm dB},
\eea
where $\lambda_{\rm dB} = 2\pi / m v$ is the de Broglie wavelength of dark matter,
$$
\lambda_{\rm dB} =  \frac{2\pi}{mv}
= 10\,{\rm AU}
\left( \frac{10^{-15}\eV}{m} \right)
\left( \frac{240\,{\rm km/sec}}{v} \right) . 
$$
As we will see in the next section, the de Broglie wavelength is a key quantity that controls the gravitational response of the wave dark matter.

We also compute the spectrum of the field fluctuations in the frequency space.  
The power and the power spectrum of the field $\hat \phi$ are defined as
\begin{equation}
P_\phi=\lim_{T\to \infty}
\frac{1}{T} \int_{-T/2}^{T/2}dt\; \hat \phi^2(t,\bfx) = \int_{-\infty}^\infty \frac{d\omega }{2\pi} S_\phi(\omega),
\end{equation}
where  the power spectral density is
\bea
S_\phi(\omega) = \lim_{T\to\infty} \frac{1}{T} |\hat \phi(\omega)|^2.
\eea
Here the Fourier component of the field is obtained via $\hat \phi(\omega) = \int_{-\infty}^{+\infty} dt \, e^{i\omega t} \hat \phi(t)$.
Since the field is a random field, the power and power spectrum are also random fields. 
The ensemble average of the spectrum is
\bea
\langle S_\phi(\omega) \rangle = 
 \frac{2\pi  \phi_0^2 }{mv} \bar{f}(v) \bigg|_{v=\sqrt{2\left(\omega/m-1\right)}}
 \label{spec}
\eea
where $\bar{f}(v)$ is given as
\bea
\bar{f}(v) = \bigg[v^2 \int d\Omega_{\hat{v}} \, f(\bfv) |\psi_\bfv(\bfx)|^2  \bigg] 
\eea
This $\bar{f}(v)$ might be interpreted as the speed distribution of wave dark matter including the gravitational focusing effect.

In above, we focus on the variance of the field itself.
Some of axion and axion-like DM searches are based on ${\cal L} \supset \partial_\mu \phi \, \bar\psi \gamma^\mu \gamma_5 \psi$.\footnote{For instance, CASPEr~\cite{Graham:2013gfa, Budker:2013hfa}, atomic magnetometers~\cite{Bloch:2019lcy, Bloch:2021vnn}, and an oscillating neutron electric dipole moment experiment~\cite{Abel:2017rtm}. }
In such cases, the signal power as well as the power spectrum would depend on the variance of the gradient of the field. 
The variance of the gradient can be obtained in the same way,
\begin{equation}\label{gradphi}
\vev{ (\hat{n}\cdot \nabla\phi)^2} = \phi_0^2 \int d^3 v\; f(\bfv) |\hat{n}\cdot \bf{\nabla}\psi_\bfv(\bfx)|^2\ .
\end{equation}
where a sensitivity axis $\hat{n}$ is a unit vector determined by the specific experimental setup. 
The spectrum can be obtained as
\bea
\vev{S_{\nabla_{\hat n}\phi}(\omega)} = \frac{2\pi \phi_0^2}{mv}
 \bigg[v^2  \int d\Omega \, f(\bfv) |\hat{n}\cdot\nabla\psi_\bfv(\bfx)|^2 \bigg]. 
\eea
Same as the power spectrum of the field, the above expression should be evaluated at $ v = \sqrt{2(\omega/m-1)}$. 

While we have mainly focused on computing variances at the same location, it is straightforward to compute correlations of fields at different locations with the same formalism. 
Such correlations contain additional information on wave DM phase at different locations, which can be used to extract further information, such as directionality of DM velocity distribution, and angular distribution of potential DM substructures, from a network of wave DM detectors~\cite{Foster:2020fln, Chen:2021bdr}.

\begin{figure}[t]
\includegraphics[width=0.23\textwidth]{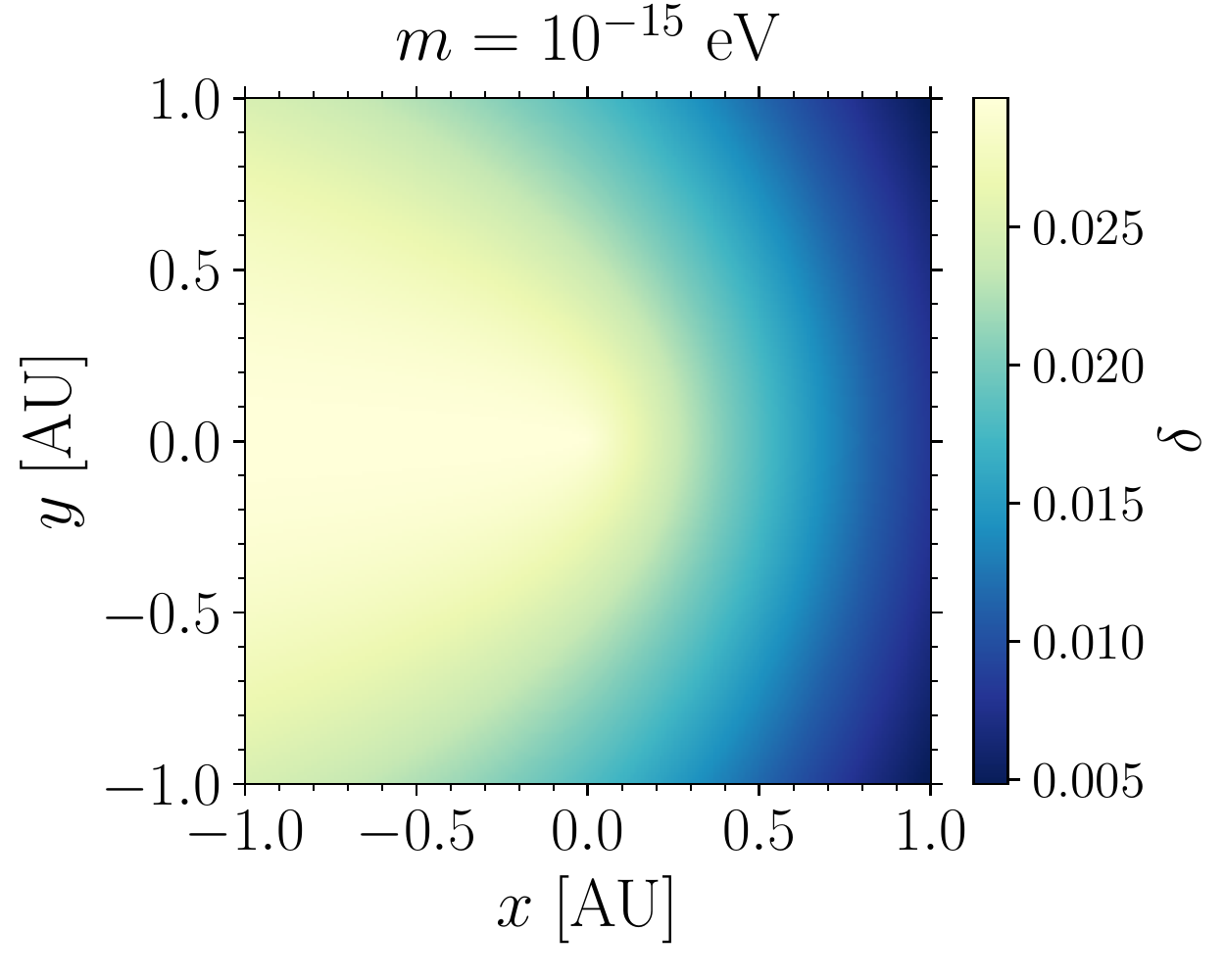}
\includegraphics[width=0.23\textwidth]{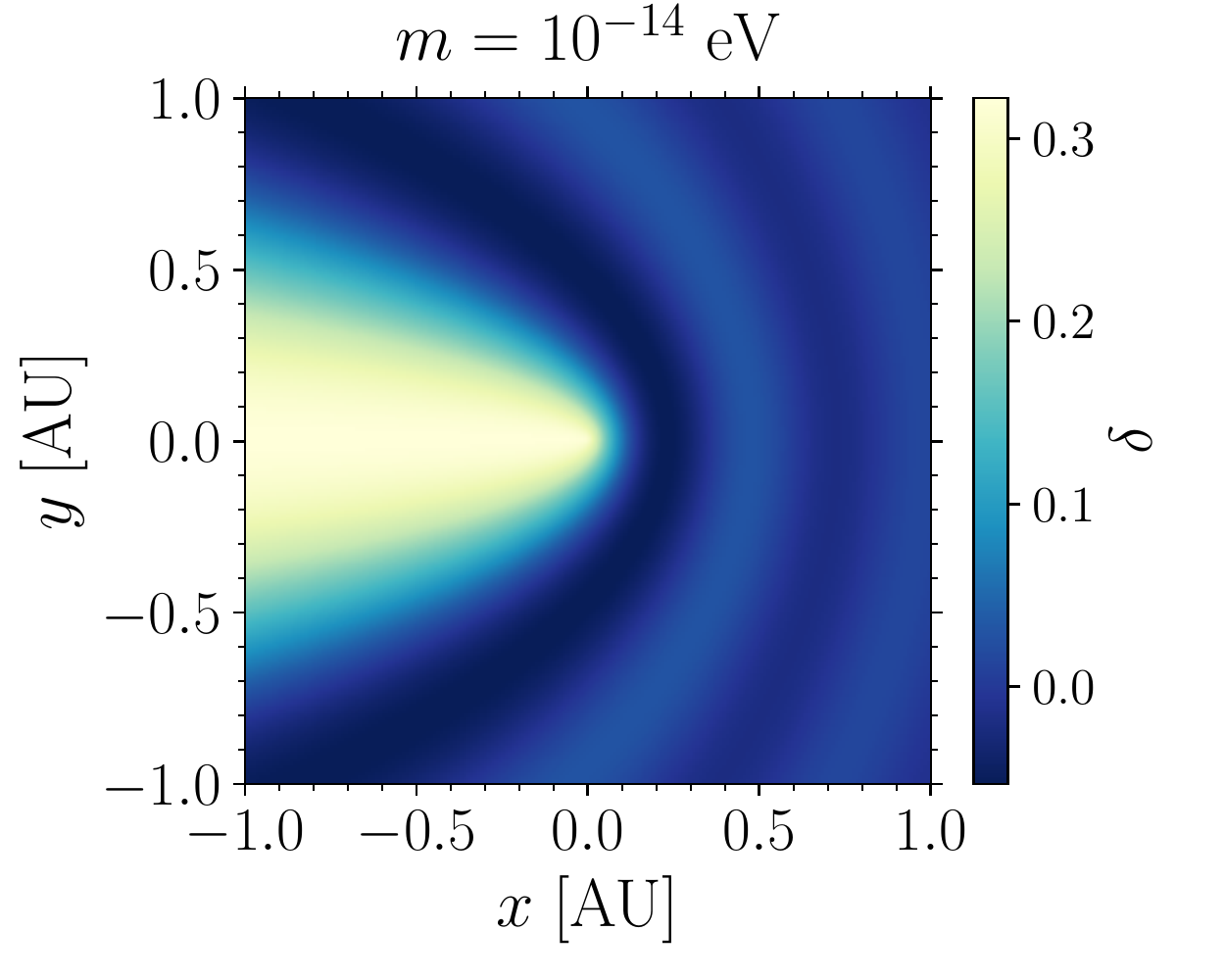}
\caption{%
We show the density contrast for the wave dark matter in the monochromatic limit, i.e. with vanishing velocity dispersion.
We choose $\bfv= (-240, \, 0, \, 0)$~km/sec and $m =10^{-15},\,10^{-14}\eV$.
The source mass is chosen to be the solar mass. 
} 
\label{fig:deltamono}
\end{figure}

\subsection{Monochromatic limit}
To examine the gravitational response of the wave dark matter, we consider the monochromatic limit. 
In this limit, the density contrast Eq.~\eqref{wave_delta} becomes
\bea
1+ \delta(\bfx) &=& |\psi_{\bfv} (\bfx)^2 | 
\nonumber\\
&=&  |\psi_{\bfv}(0)|^2 |M[ i \beta, 1, i m v r (1-\hat{v} \cdot \hat{x})]|^2 . 
\eea
Here $\psi_{\bfv}(0)$ is the wave function at the origin. 
The squared amplitude at the origin is
\bea
|\psi_{\bfv}(0)|^2 = \frac{2 \pi \beta}{1-e^{- 2 \pi \beta} } .
\label{delta_mono_max}
\eea
The above is identical to the Sommerfeld enhancement for dark matter annihilation when the interaction is mediated by the Coulomb interaction~\cite{Arkani-Hamed:2008hhe}.

\begin{figure}[t]
\includegraphics[width=0.4\textwidth]{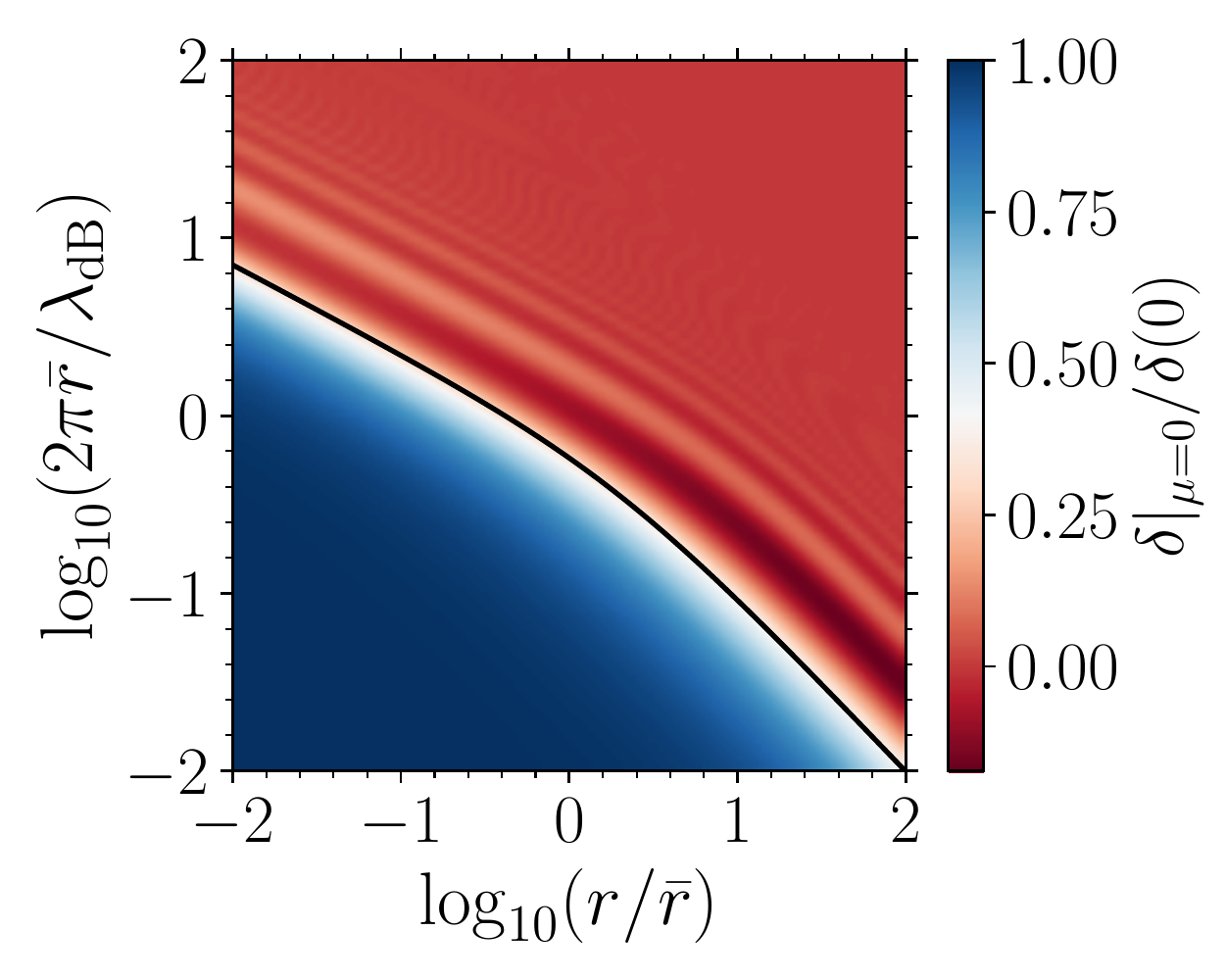}
\caption{The density contrast at $\hat{x}\cdot\hat{v} = 0$. 
The constant density contrast is observed for small radii satisfying $m\tilde{v} r <1$.
The black line corresponds to $m\tilde v r =1$. }
\label{fig:wave_mono_1}
\end{figure}

\begin{figure}[t]
\includegraphics[width=0.4\textwidth]{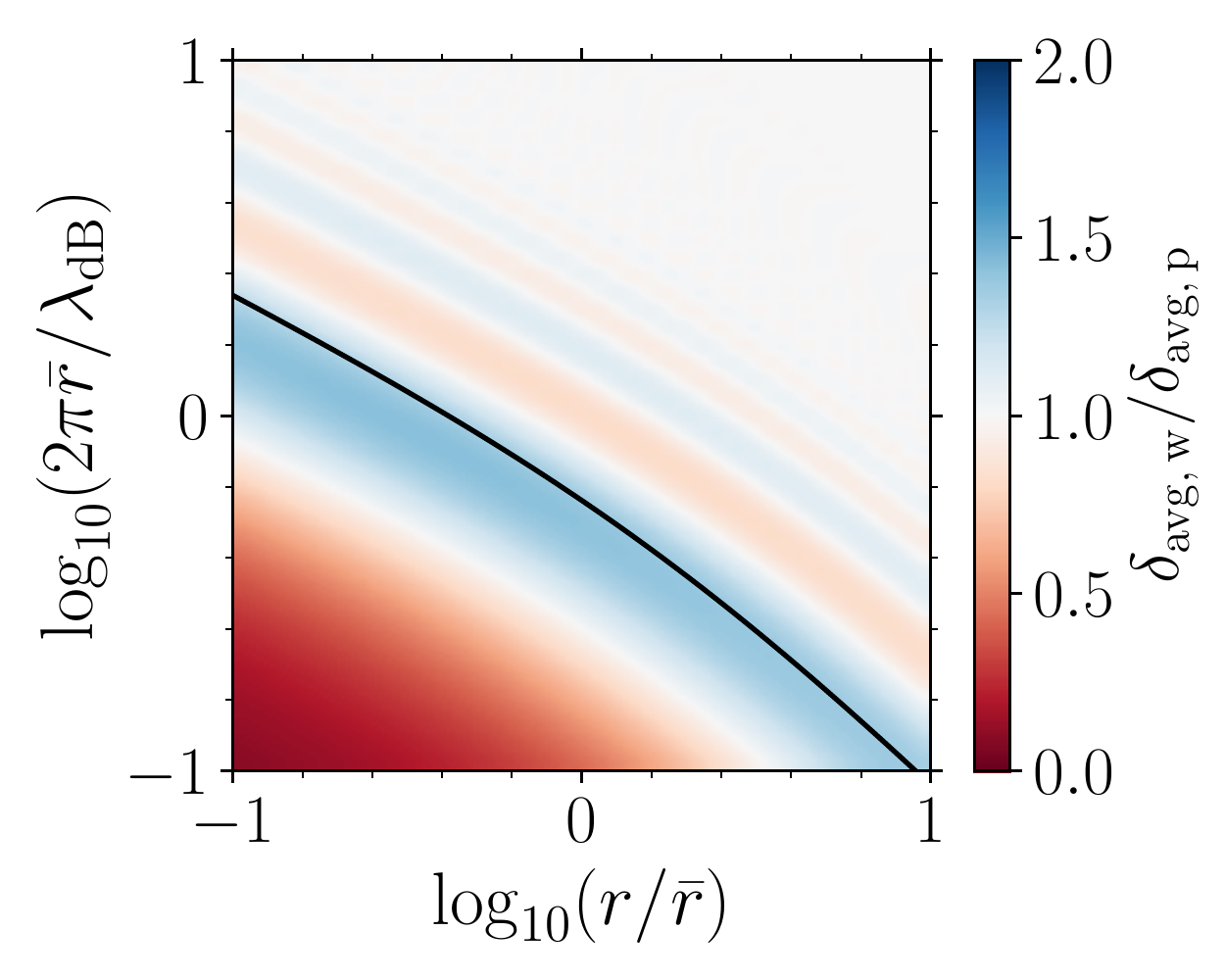}
\caption{The ratio between the density contrast averaged over the solid angle for wave and particle dark matter. 
For large radii, $m\tilde v r \gtrsim 1$, the averaged density contrast for wave dark matter approaches that of the particle dark matter. 
The solid black line represents $m\tilde v r = 1$.}
\label{fig:wave_mono_2}
\end{figure}

In Figure~\ref{fig:deltamono}, we show the density contrast of wave dark matter in the monochromatic limit.
We choose $\bfv = (- 240,\, 0, \, 0)$~km/s, $m=10^{-15}$, $10^{-14}$~eV, and $M = M_\odot$. 
We observe a similar density wake field along the direction of dark matter velocity.

Some of wave properties in the density wake are revealed upon a closer examination of Figures~\ref{fig:deltamono}--\ref{fig:wave_mono_2}.
Contrary to the particle dark matter, the density contrast saturates to a constant value, $\delta\simeq |\psi_{\bfv}(0)|^2 -1$, for sufficiently small $r$. 
In addition, there is an oscillating pattern of density contrast over a certain specific spatial scale.
Both of wave characteristics are controlled by de Broglie wavelength defined as
\bea
\tilde \lambda_{\rm dB} (r) = \frac{2\pi}{m\tilde v (r)}
\eea
where $\tilde v$ is the speed at $r$, given as
\bea
\tilde v^2(r) = v^2 + v_e^2(r).
\eea
Note that we have used the speed at a radial distance $r$, rather than the velocity asymptotically far away from the source $M$.

To illustrate how this de Broglie wavelength controls the monochromatic result, we show in Figure~\ref{fig:wave_mono_1} the density contrast normalized with the central value, $\delta(\bfx)/ \delta(0)$, as a function of two parameters, $r/\bar r$ and $2\pi \bar{r} /\lambda_{\rm dB}$. 
Here $\lambda_{\rm dB} = 2\pi / m v$ where $v$ is the dark matter velocity far away from the source. 
The black line in the figure is $\tilde \lambda_{\rm dB} = 2\pi r$, i.e. $m\tilde{v}(r) r =1$. 
We observe that the density contrast saturated to $\delta(0)$ for sufficiently small r, $m \tilde v(r) r \lesssim 1$. 
For $m \tilde v(r) r \gtrsim 1$, the wave density contrast oscillates rapidly but on average, it behaves similar to that of particle dark matter.

In Figure~\ref{fig:wave_mono_2}, we plot the ratio of density contrast averaged over the solid angle for the wave and particle dark matter. 
For particle dark matter, we choose $1+\delta_{\rm avg,\, p} = \sqrt{1 + v_e^2 / v^2}$. 
We observe that for $m\tilde v(r) r \gtrsim1$, the averaged density contrast for the wave dark matter becomes increasingly similar to the particle result, while for $m\tilde v(r) r \lesssim 1$, the averaged density contrast is suppressed as $1+\delta_{\rm avg} \approx \pi ( 2\pi r/\tilde \lambda_{\rm dB} ) (1 + v_e^2 / v^2 )^{1/2}$.

In both figures, we observe that a key parameter controlling the wave dark matter density contrast is the relative size of the de Broglie wavelength $\tilde\lambda_{\rm dB}$ and $r$. 
For $m\tilde v r \gtrsim 1$, the response of wave dark matter to the gravitational potential is similar to that of particle dark matter, while for $m\tilde v r \lesssim 1$, wave characteristics appear. 
The limit $m \tilde v r \gg 1$ is the limit at which the wave dark matter approaches to particle dark matter. 
This will be discussed in more detail in Section~\ref{sec:classical_limit} and in Appendix~\ref{app:semiclassical}.

\subsection{Comparison}
The monochromatic results clearly show the role of de Broglie wavelength.
We now consider the density contrast of wave dark matter with velocities distributed according to the Maxwell-Boltzmann distribution to compare the wave dark matter with particle dark matter on an equal footing.

\begin{figure}
\centering
\includegraphics[width=0.45\textwidth]{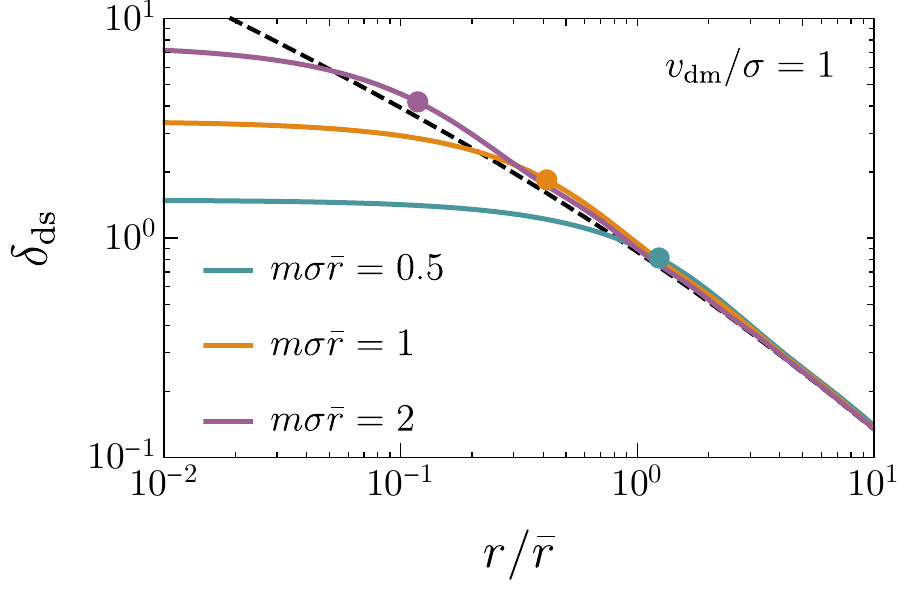}
\includegraphics[width=0.45\textwidth]{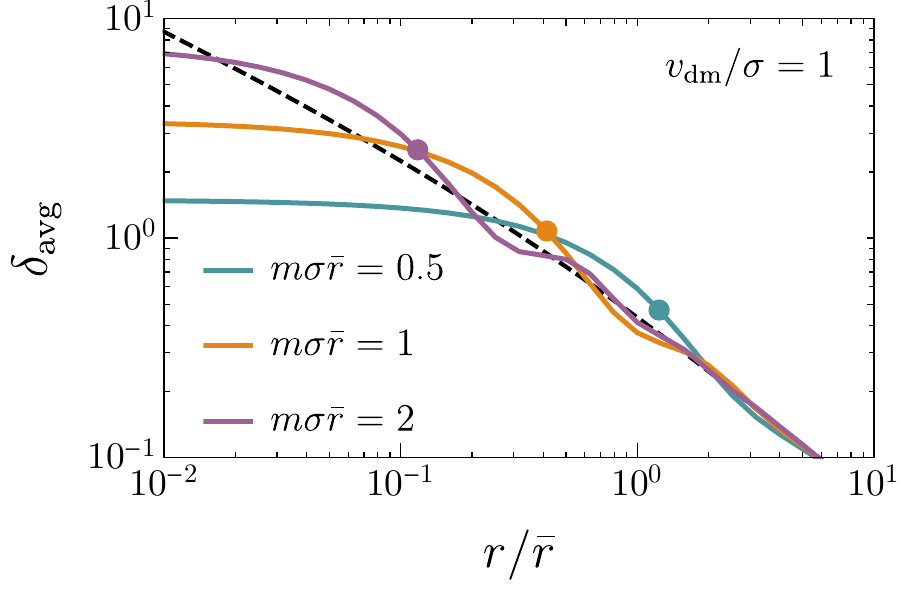}
\caption{(Top) We show the density contrast along the downstream. 
Each color represents different values of $m\sigma \bar r$.
The circle represents the point $m\tilde v(r) r =1$.
The black dashed line is the result of particle dark matter obtained in Section~\ref{sec:focusing_particle}. 
We observe that the density contrast approaches the particle dark matter result for $m\tilde v r \gtrsim 1$, while it saturates to a constant value for $m \tilde v r \lesssim 1$. 
(Bottom) We show the averaged density contrast. For both figures, we choose $v_{\rm dm}/\sigma =1$. }
\label{fig:compare}
\end{figure}

In Figure~\ref{fig:compare}, we present two quantities, the density contrast along the downstream (top) and the density contrast averaged over the solid angle (bottom). 
For both plots, we choose $v_{\rm dm}/ \sigma =1$. 
As in the monochromatic case, particle-to-wave transition takes place when $m\tilde v r \simeq 1$.
Since we consider a medium with nonvanishing velocity dispersion, we should be careful interpreting $\tilde v^2 = v^2 +v_e^2$ since the final density contrast is the result of the superposition of different velocity components. 
For the following discussion, we redefine $\tilde v^2 = v_{\rm dm}^2 + \sigma^2 + v_e^2$ for the medium with nonzero velocity dispersion. 
In the figure, the black dashed line represents the result of particle dark matter, while colored lines represent the result of wave dark matter with $m \sigma \bar r = 0.5$ (cyan), $1$ (orange), $ 2$ (purple). 
For $m \tilde v  r \lesssim 1$, the density contrast saturates to
\bea
1 + \delta(0) =\int d^3v \, f(\bfv)  |\psi_{\bfv}(0)|^2 
\approx \frac{2\pi \langle \beta \rangle}{1 - e^{-2\pi \langle \beta \rangle}}
\eea
with $\langle \beta \rangle = (GMm/v_{\rm dm}) \,{\rm erf}(v_{\rm dm}/2\sigma)$.
For $m \tilde v r \gtrsim 1$, the wave result converges to the particle dark matter result.

\section{Dark matter in solar system}\label{sec:application}
We apply our previous discussion to dark matter structures in the solar system. 
We work in the Galactic coordinate system where the origin of coordinates is at the location of the Sun. 
In the rectangular Cartesian coordinate system, each component of $(X, Y, Z)$ points towards the Galactic center, the direction of Galactic rotation, and the Galactic north pole. 
In this coordinate system, we compute the density contrast at the position of Earth in different times in a year for dark matter substructures with different kinematic properties. 

For the velocity distribution, we use the Maxwell-Boltzmann distribution, Eq~\eqref{mb}, for most of dark matter substructures. The distribution is therefore characterized by the mean velocity $\langle \bfv \rangle = \bfv_{\rm dm}$ and the one-dimensional velocity dispersion $\sigma$.

\subsection{Halo dark matter}

We begin with the relaxed halo dark matter component. 
In the Galactic coordinate, the halo dark matter can be represented by the standard halo model with a mean velocity in the Galactic coordinate~\cite{2010MNRAS.403.1829S, 2019ApJ...871..120E},
\bea
\bfv_{\rm dm} = - \bfv_\odot = - (11,\, 241, \, 7)\,{\rm km/sec}
\eea
The velocity dispersion is
\bea
\sigma = v_c (R_\odot) / \sqrt{2} = 162\,{\rm km/sec}
\eea
where $v_c (R_\odot) \simeq 229\,{\rm km/sec}$ is the circular velocity of Milky Way at the position of the Sun~\cite{2019ApJ...871..120E}. 

%% HALO FIG %%
\begin{figure}[t]
\centering
\includegraphics[width=0.5\textwidth]{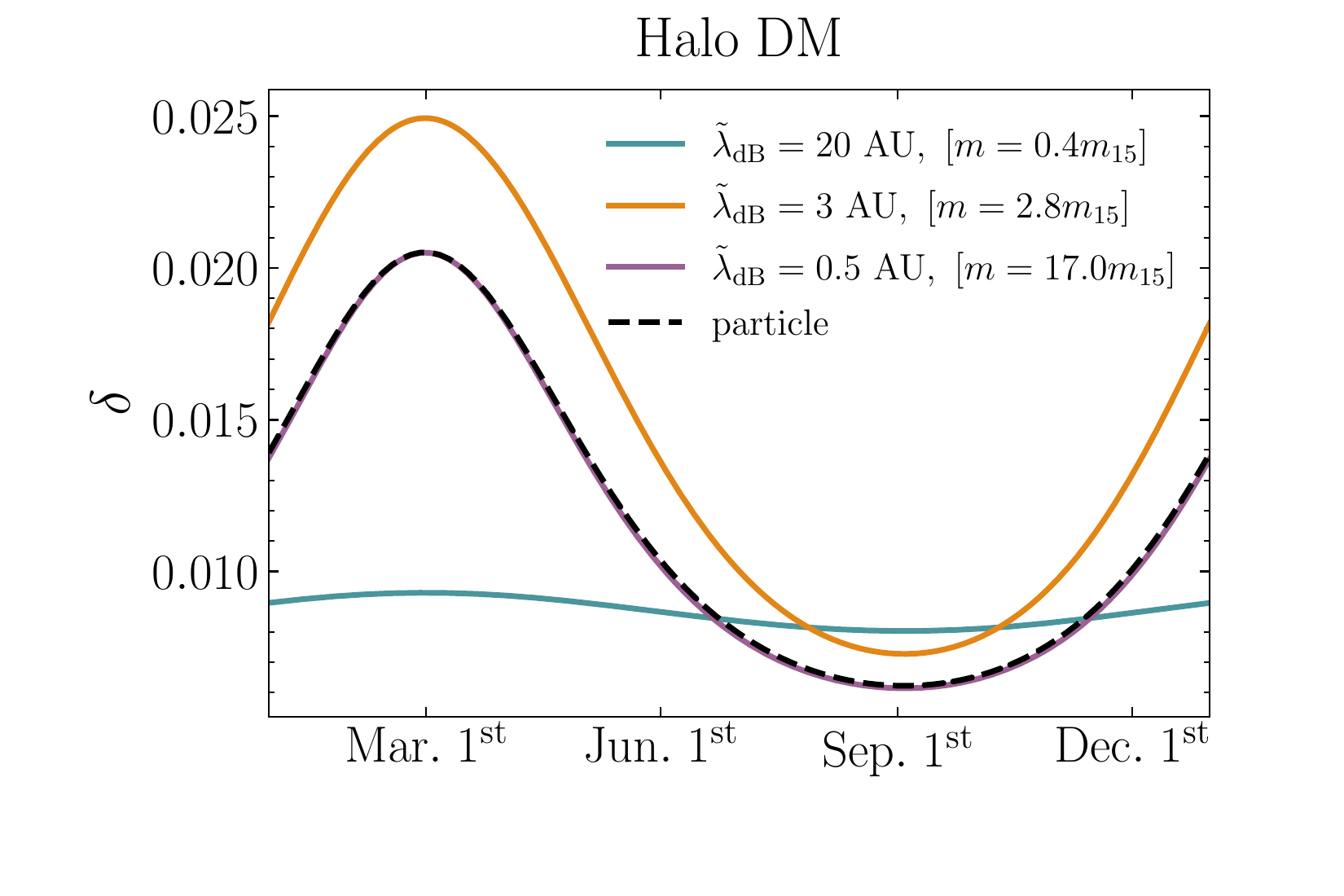}
\caption{Density contrast for halo dark matter at the position of Earth during a year 2021. 
The black dashed line is for the particle dark matter, while each colored line represents the wave dark matter result for $\tilde{\lambda}_{\rm dB}  = 20 , \, 3, \,  0.5$ AU. 
Corresponding wave DM mass with respect to $m_{15}=10^{-15}\eV$ is also shown. 
The density contrast flatten for $m \tilde v r \lesssim 1$, while it approaches the particle dark matte result for $m\tilde v r \gtrsim 1$ similarly to the halo dark matter example. 
Around $m\tilde{v} r \sim 1$, the wave density contrast becomes $\sim 25~\%$ larger than that of particle dark matter. }
\label{fig:halo}
\end{figure}

In Figure~\ref{fig:halo}, we show the density contrast of wave dark matter at the position of the Earth throughout the year of 2021 for different values of $m \tilde v r$. 
Since $v_e \simeq 42\,{\rm km/sec} \ll \sigma, v_{\rm dm}$, we expect that the overall amplitude of density contrast is $\delta \simeq v_e^2 /2 \sigma^2 \simeq 0.03$ at least for the particle dark matter, which is confirmed by the black dashed line in the figure. 
The maximum of the density contrast takes place around March 1st since $\mu = \hat{x}\cdot\hat{v}_{\rm dm}$ takes the largest value over a year. 
This pattern can be seen also from the wave dark matter density contrast with varying degree of annual modulation strength. 
While the maximum and minimum arise at the same time, we observe that the density contrast flattens for $m \tilde v r \lesssim 1$ as the de Broglie wavelength becomes larger than an astronomical unit.
For $m \tilde v r \gg 1$, the result can be approximated as that of particle dark matter density contrast. 
We also observe that the wave density contrast is enhanced by $\sim 25\%$ compared to the particle density contrast at $m \tilde v  r \sim 1$.

\begin{figure}[t]
\centering
\includegraphics[width=0.5\textwidth]{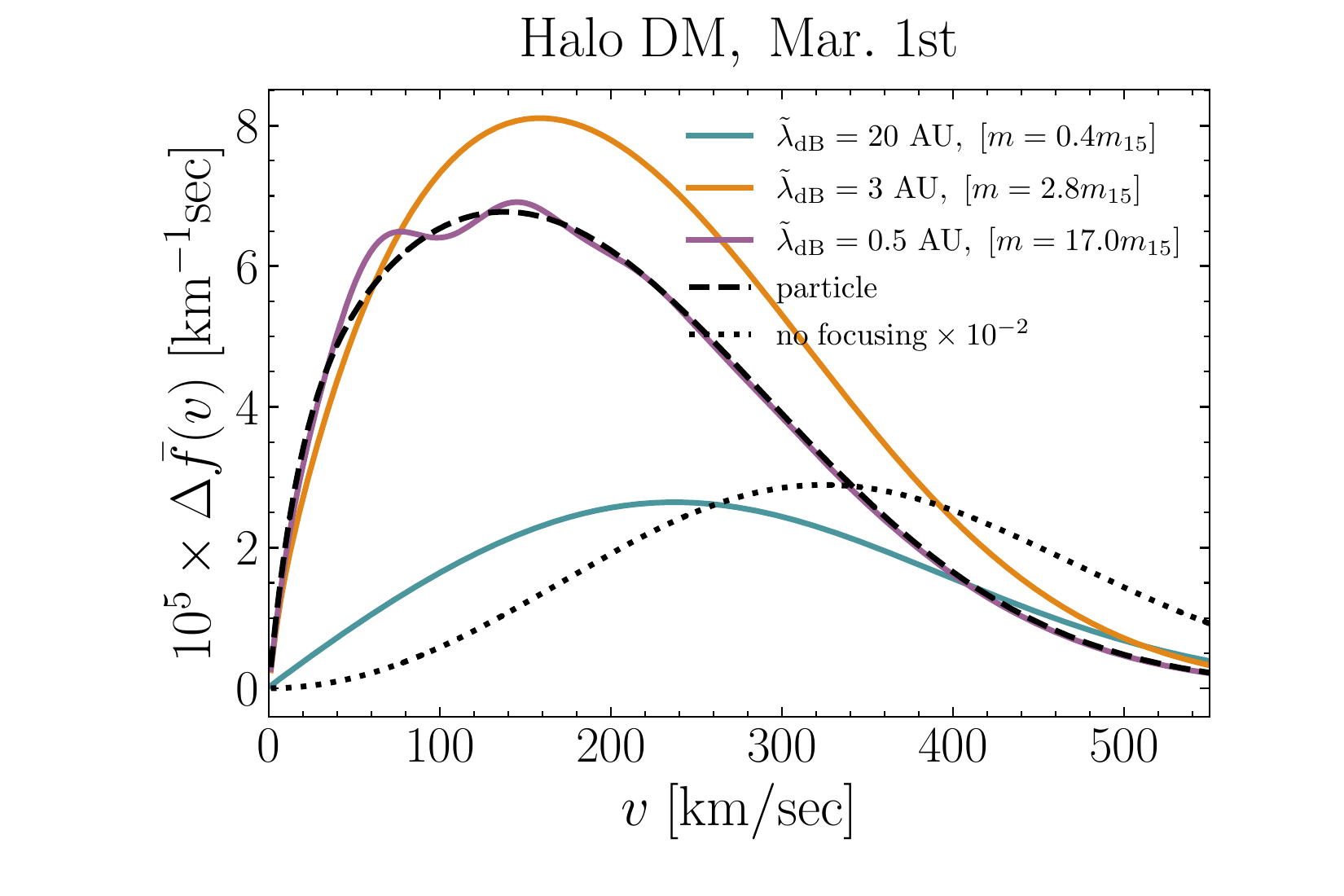}
\includegraphics[width=0.5\textwidth]{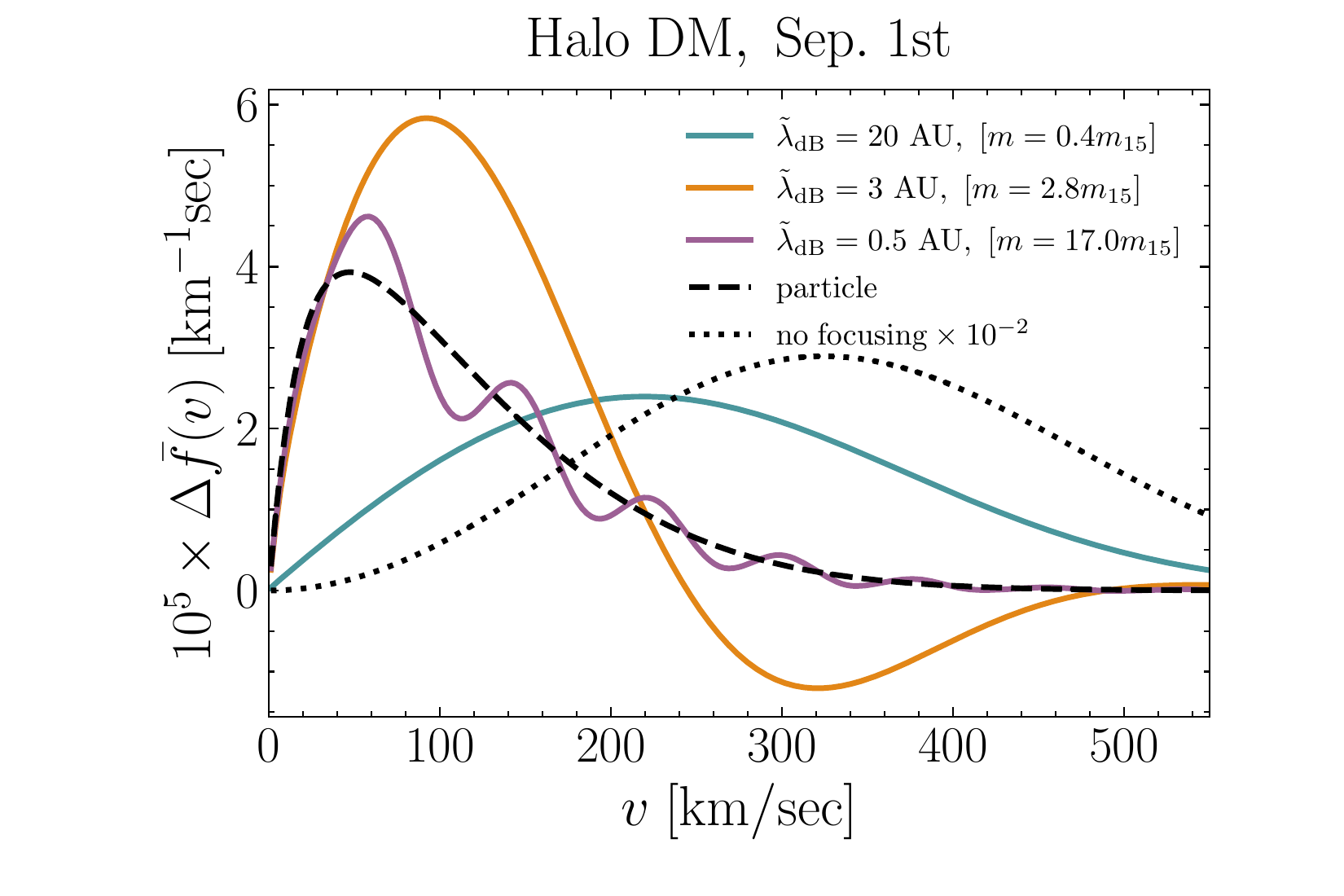}
\caption{(Top) The difference of speed distribution with and without the gravitational focusing, computed at March 1st, 2021. (Bottom) Same as top panel but computed at September 1st, 2021. 
Each colored line represents $\tilde{\lambda}_{\rm dB} = 20, \,3 , \, 0.5$ AU.
The black dashed line is the same result for the particle dark matter, and the black dotted line is the rescaled speed distribution without gravitational focusing. }
\label{fig:halo_speed}
\end{figure}

In Figure~\ref{fig:halo_speed}, we also show the modification of the speed distribution for the wave dark matter.
More specifically, we plot 
\bea
\Delta \bar{f}(v) = v^2 \int d\Omega_{\hat v} f(\bfv) ( | \psi_{\bfv}(\bfx)|^2 -1 ),
\eea
which is the difference between the speed distribution with and without the gravitational focusing. 
We choose March 1st and September 1st for the plots. 
As it is clear from the figures, the effect of gravitational focusing is prominent in the low velocity tail. 
Each colored line represent the result of wave dark matter for different values of $m\tilde v r$.
The black dashed line is the modification of speed distribution in the case of particle dark matter.
The dotted line is rescaled speed distribution without gravitational focusing. 

The variance of the projection of the field gradient along the sensitivity axis of a detector for the halo DM component is shown in Appendix \ref{app:gradient}.

\subsection{Gaia Sausage}
We consider a dark matter substructure with anisotropic velocity distribution. 
Such a dark matter substructure is motivated by the observation of a stellar  population with an anisotropic velocity distribution, referred to as Gaia-Sausage or Gaia-Enceladus~\cite{2018MNRAS.478..611B, 2018Natur.563...85H}. 
It is inferred that this stellar component originates from a relatively recent merger with a luminous satellite galaxy of mass $M > 10^{10}M_\odot$. 
It is naturally expected that not only stars but also dark matter is accreted on to the Milky Way from the same event. 
Studies have shown that the kinematic properties of accreted stars and dark matter are similar~\cite{Necib:2018iwb, Necib:2018igl} and that it could consitute ${\cal O}(10\%)$ of local dark matter~\cite{Evans:2018bqy}.

We inspect the response of sausage-component dark matter to the solar gravitational potential.
Since the velocity structure is anisotropic, we use a general three-dimensional normal distribution for the DM velocity distribution,
\bea
f(\bfv) &=& \frac{1}{(2\pi)^{3/2} \sqrt{\det \Sigma} } 
\nonumber\\
&&
\times
\exp
\left[
- \frac{1}{2} ( \bfv - \bfv_{\rm dm} ) \cdot \Sigma^{-1} \cdot (\bfv - \bfv_{\rm dm } )
\right]
\eea
where $\Sigma$ is the covariance matrix. 
We parameterize sausage dark matter with $\bfv_{\rm dm} = - \bfv_\odot$
and
\bea
\sigma_r = 256\,{\rm km/sec},
\quad
\sigma_\theta = \sigma_\phi =81 \,{\rm km/sec},
\eea
where $\Sigma = {\rm diag}(\sigma_r^2 , \sigma_\theta^2, \sigma_\phi^2)$~\cite{OHare:2019qxc}. 
This covariance matrix reflects the high velocity anisotropy of this substructure. 

%% Sausage Figure %%
\begin{figure}[t]
\centering
\includegraphics[width=0.5\textwidth]{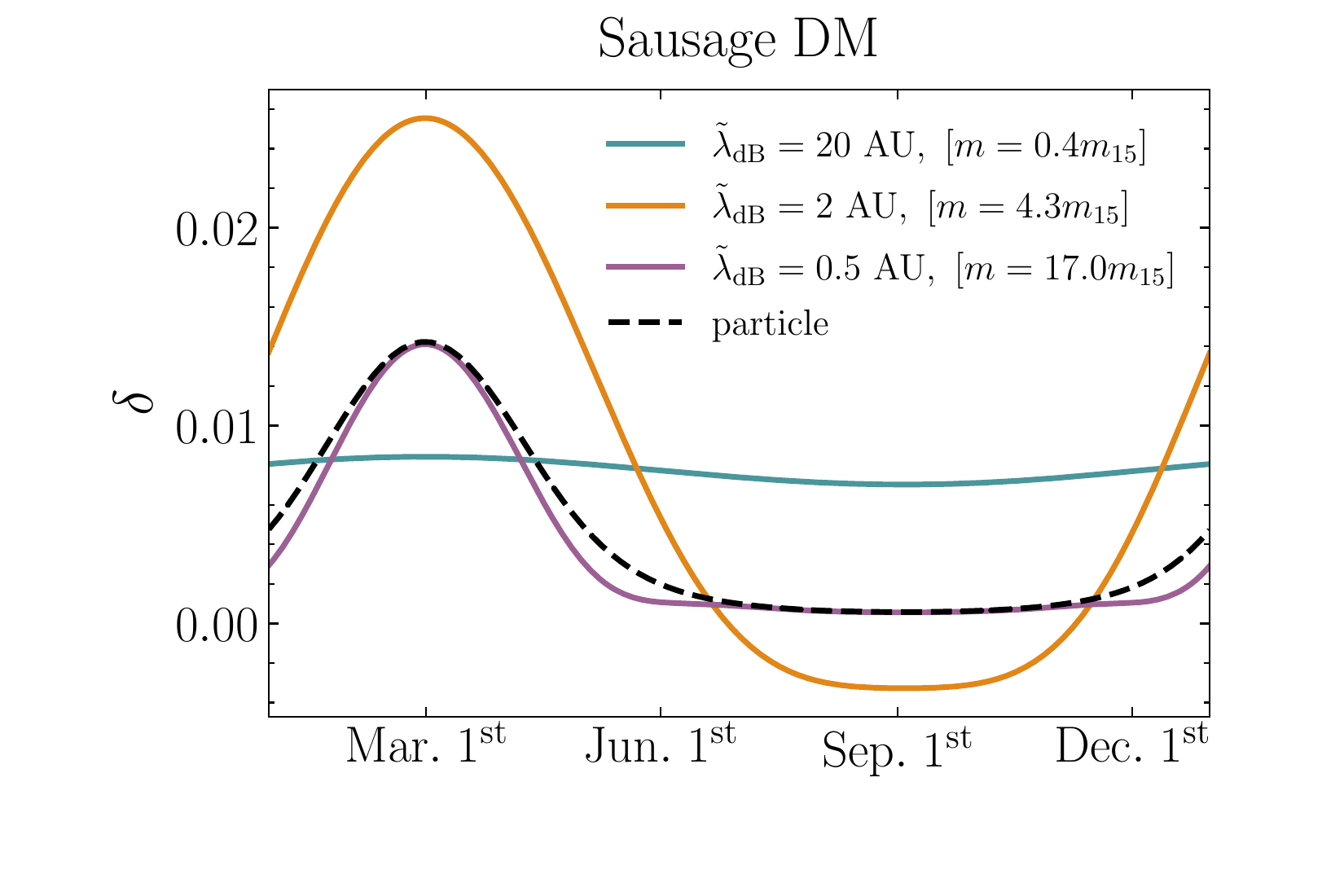}
\caption{
Same as Figure~\ref{fig:halo} but for sausage dark matter component.
Around $\tilde{\lambda}_{\rm dB}\simeq 2$ AU, corresponding to $m \tilde v r \simeq 3$, the wave dark matter contrast can be almost a factor two larger than the particle density contrast.
We also observe a slight underdensity around fall. }
\label{fig:sausage}
\end{figure}

In Figure~\ref{fig:sausage}, we present the density contrast for the sausage dark matter component. 
We observe a similar pattern as in the example of halo dark matter. 
Overall amplitude is similar to the halo dark matter component, despite that the velocity distribution is highly anisotropic.
The maximum and the minimum of the density contrast also take place at the same period of the time. 
A slight underdensity is also observed around fall.

\subsection{Dark disk}
As a next example, we consider dark matter in the form of disk.
This dark matter substructure is motivated by the thick stellar disk. 
To explain the vertical structure of stellar disk, several scenarios have been suggested.
For instance, thick stellar disk is composed of stars that are accreted from satellite galaxies~\cite{2003ApJ...597...21A} or stars in pre-existing thin disk are heated through merger events~\cite{2008ApJ...688..254K, 2008MNRAS.391.1806V}. 
It has been suggested that such merger events could naturally lead to the formation of thick dark disk as the accreted dark matter is dragged to the disk plane through dynamical friction~\cite{2008MNRAS.389.1041R}.
The resulting dark matter disk is co-rotating with Galactic disk with slightly smaller circular velocity by $\Delta v = 50$~km/sec. 
While it is still not clear whether the aforementioned merger scenario can solely explain the evolution of stellar disk (see a review~\cite{2013A&ARv..21...61R} and references therein), it is still worthwhile to investigate the response of such dark matter component to the solar gravitational potential.

\begin{figure}[t]
\centering
\includegraphics[width=0.5\textwidth]{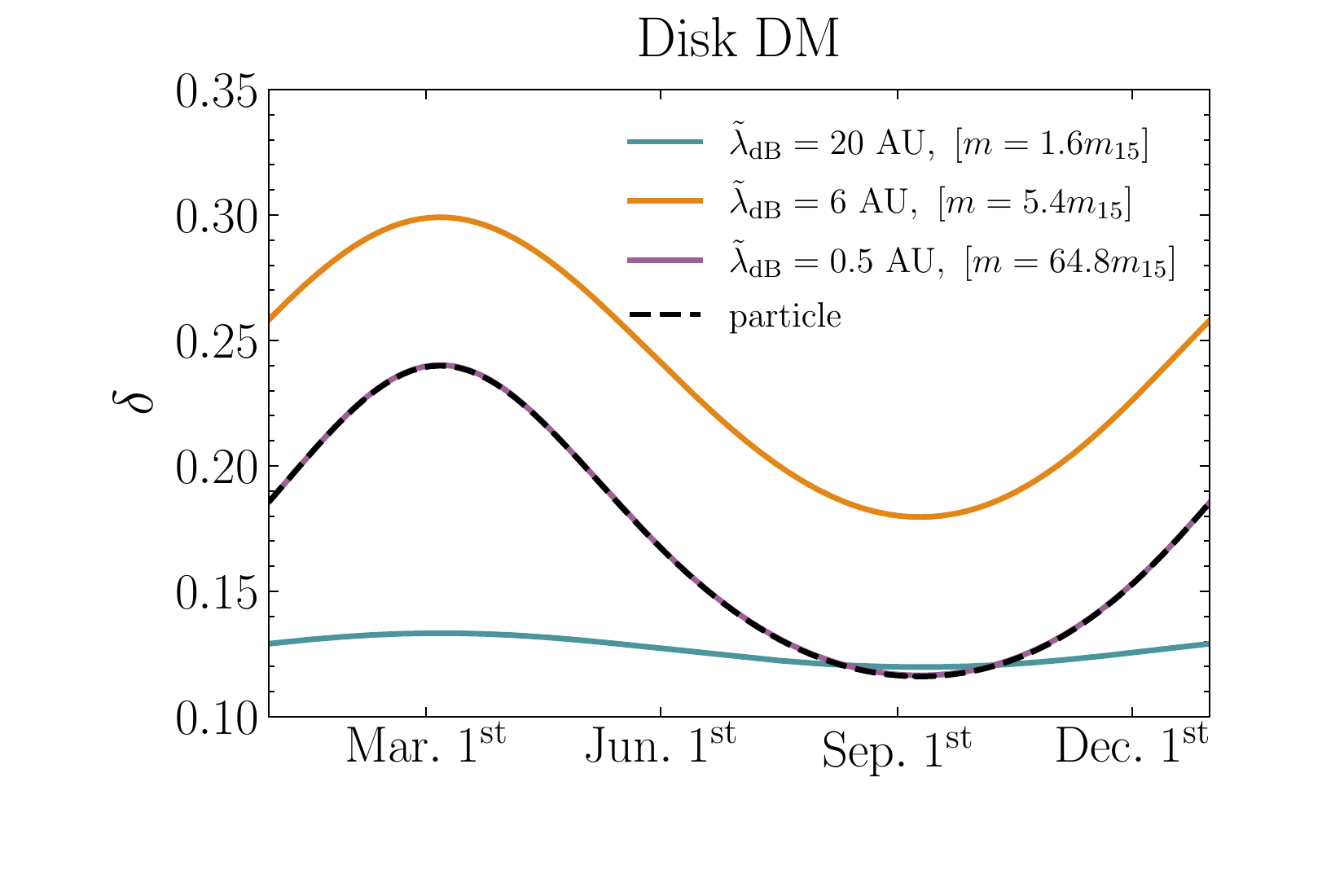}
\caption{%
Same as Figures~\ref{fig:halo} and \ref{fig:sausage}, but for the dark disk. 
Due to the small mean velocity and dispersion, the overall amplitude of density contrast is larger than the previous DM structures. 
}
\label{fig:disk}
\end{figure}

For the investigation of dark disk, we parameterize the mea velocity and velocity dispersion as
$$
\bfv_{\rm dm} = (0 , - 50, 0 )~{\rm km/sec}
\quad
\textrm{and}
\quad
\sigma = 50 \, {\rm km/sec}.
$$
While we refer this structure to dark disk, it could also represent any cold dark matter substructure with a small mean velocity.\footnote{The velocity dispersion of thick stellar disk is $(\sigma_r,\sigma_\theta,\sigma_\phi) \simeq (63, 39, 39)$~km/sec~\cite{2003A&A...398..141S}, but we assume an isotropic distribution for dark matter for simplicity. 
In addition, even in the merger scenario, the dark matter kinematic structure depends on the merger details. Numerical studies have shown that the DM kinematic structure follows that of accreted stars, and the accreted stars are hotter than heated thin disk stars, indicating that the velocity dispersion of dark matter is likely to be higher than that of stellar thick disk~\cite{2008MNRAS.389.1041R, 2009ApJ...703.2275P}. 
Nonetheless, we choose $\sigma =50$~km/sec as this benchmark more clearly shows the characteristic behavior of gravitational focusing.}

In Figure~\ref{fig:disk}, we show the density contrast for the dark disk. 
The pattern of the density contrast is similar to previous cases, but with a larger amplitude due to the small velocity and dispersion. 
In this particular example, the disk dark matter can be focused and its density near the orbital trajectory of Earth could be $\sim 30\%$ larger than its density far away from the Sun.

The variance of the field gradient in the case of a dark disk component is shown in Appendix \ref{app:gradient}.

\subsection{Stream}
As a last example, we consider dark matter streams, whose kinematic structure is similar to that of stellar stream. 
Stellar streams are stellar substructures and they are coherent both spatially and kinematically. 
If stellar streams originate from dwarf galaxies, it is likely that there are associated dark matter streams with similar kinematic properties~\cite{Necib:2018igl}. 
In the rest frame of Milky Way dark matter halo, streams typically have large streaming velocity with a small velocity dispersion. 
Various streams and substructures have been observed in the inner halo in the past decades. See a review~\cite{2020ARA&A..58..205H} for details on stellar streams and substructures in Milky Way.

To investigate the gravitational response of fast-moving cold objects like streams, we consider 
$$
v_{\rm dm} =400~{\rm km/sec},
\quad
\sigma =30\, {\rm km/sec }. 
$$
Note that the mean speed in the rest frame of the Sun might be smaller than the above value for prograde streams.
In such cases, the density contrast as well as the spectrum would resemble that of dark disk.  
For the direction of the stream, we consider two different inclination angles with respect to the ecliptic plane, $\theta_{\rm dm} = 0, \, \pi/6$.

\begin{figure}[t]
\includegraphics[width=0.5\textwidth]{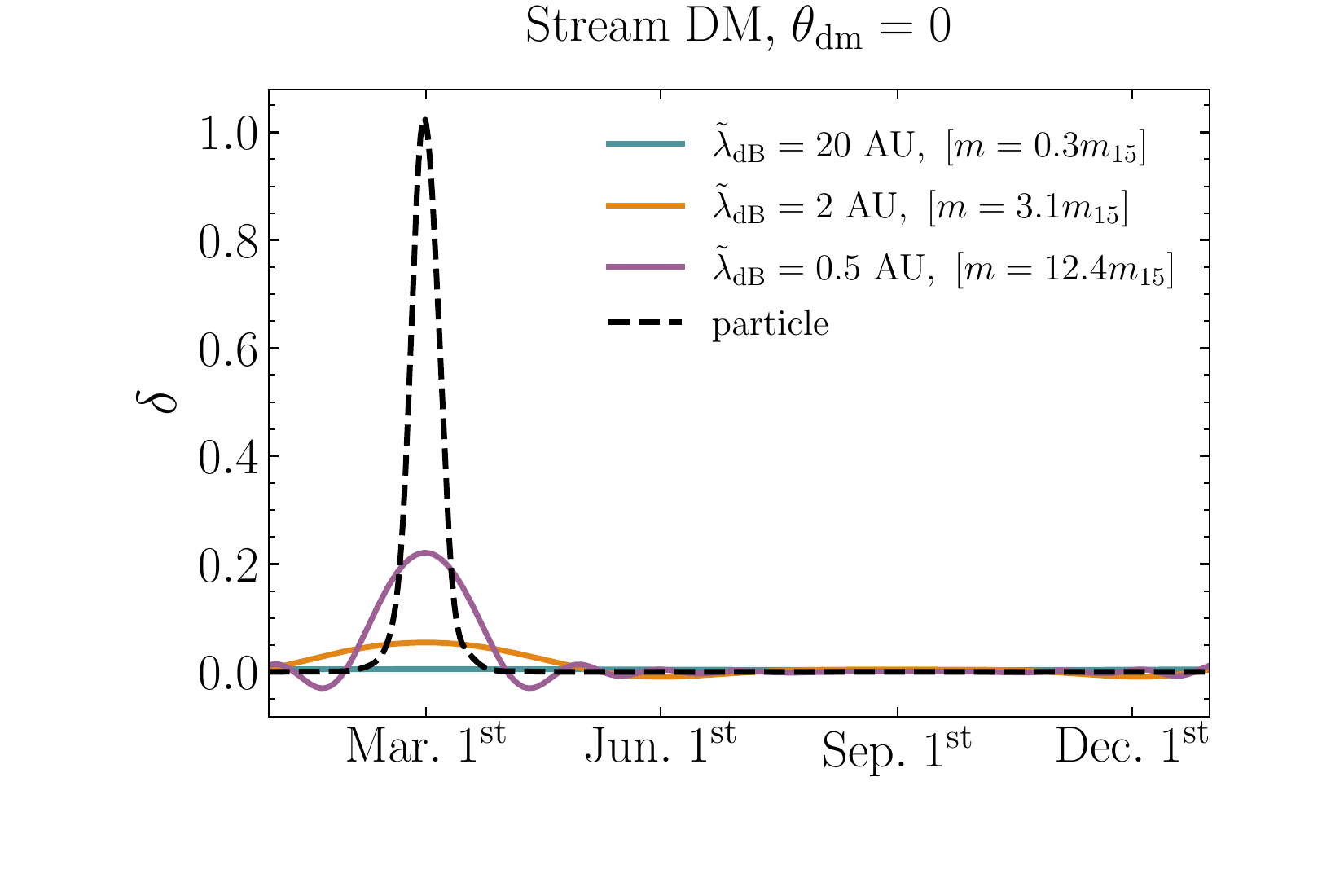} 
\includegraphics[width=0.5\textwidth]{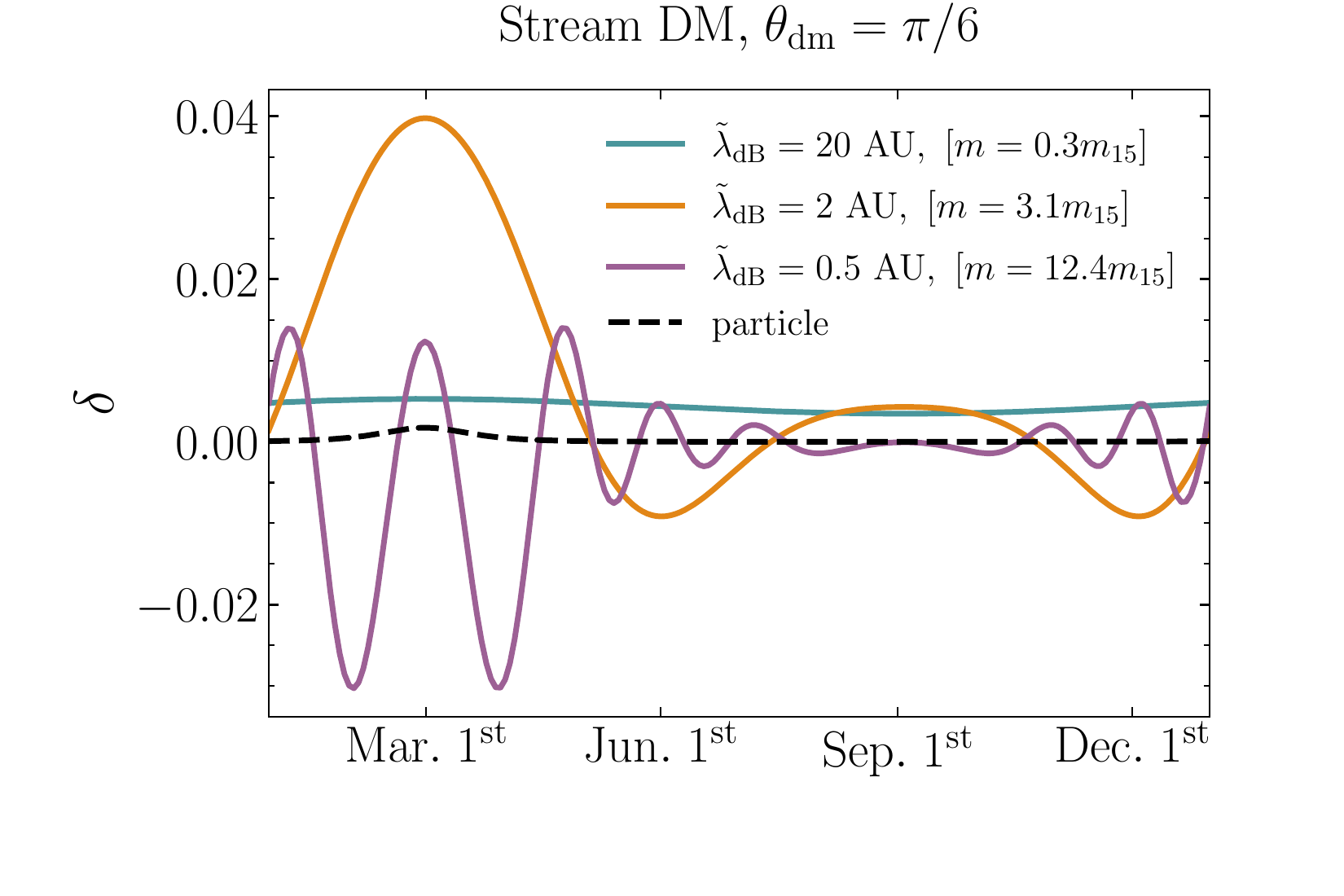} \caption{(Top) The density contrast for the stream for the vanishing inclination angle. 
Due to the hierarchy between the mean velocity and the dispersion, the density contrast is highly localized near March 1st.
The wave dark matter contrast for $\tilde{\lambda}_{\rm dB}=0.5$ AU, corresponding to $ m\tilde v r = 10$, is still quite different from the particle dark matter result. See the main text for the explanation. 
(Bottom) The density contrast for the stream for the inclination angle $\theta_{\rm dm} = \pi/6$. 
A small velocity dispersion for stream does not completely erase the oscillations of the wave function, and therefore, much faster oscillation of density contrast is obtained in this example. }
\label{fig:stream}
\end{figure}

In Figure~\ref{fig:stream}, we show that the density contrast for dark matter streams for two different inclination angles. 
For the case of vanishing inclination angle $\theta_{\rm dm} = 0$, we observe interesting differences between the particle dark matter and wave dark matter.
Even for $m\tilde v r  = 10$, the wave contrast is much broader than the particle contrast in the time-axis. 
We have argued that the wave dark matter contrast approaches to the particle contrast for $m \tilde v r \gg 1$, based on quantities like $\delta_{\rm avg}$, but note that $\delta_{\rm avg}$ is the averaged over the solid angle. 
At the same time, we have seen both for the particle and wave dark matter that there exists a certain angular scale $\Delta \mu$ over which the density contrast takes nonvanishing value. 
For the particle dark matter, this angular scale is typically $\Delta \mu \sim (\sigma / v_{\rm dm})^2$ because of the non-vanishing velocity dispersion, while for the wave dark matter, due to the de Broglie wavelength, it is $\Delta \mu \sim \max[ 1 / m \tilde v r, (\sigma/v_{\rm dm})^2 ]$. 
The stream that we consider has a large hierarchy between $v_{\rm dm}$ and $\sigma$, and hence, the particle density contrast is nonvanishing only for a relatively small period of time in a year, i.e. $\Delta t \sim (\sigma/v_{\rm dm})^2 {\rm yr}$, while for the wave dark matter, for $m\tilde v r =10$, the angular spread $\Delta \mu$ is still controlled by $1 / m \tilde v r$ rather than $(\sigma/  v_{\rm dm})^2$. 
Therefore, the wave contrast in Figure~\ref{fig:stream} is less localized in time (or in space) compared to the particle contrast. 

This same feature can also be observed in the example of $\theta_{\rm dm} = \pi/6$. 
In this case, the density contrast for the particle dark matter takes nearly vanishing value, while for the wave dark matter, it takes still non-vanishing value at the level of a few percent. 
Another interesting feature is that the pattern of density contrast over a year yields much faster oscillations, which is clearly distinct from other substructures.
This can be attributed to the small velocity dispersion; the wave behaves more a like monochromatic wave and the characteristic oscillation patterns survive even after the waves with different wavenumber are mixed together.

\section{Discussion}\label{sec:discussion}

\subsection{Coordinate transformation}
We have worked in the Galactic coordinate system.
This coordinate is useful for the investigation of the effects of gravitational focusing since the gravitational focusing effects can be easily isolated. 
Since terrestrial experiments are bound to the Earth, it is worthwhile to mention how the observables such as the density contrast and spectrum change in the detector rest frame. 

The density contrast does not change at all. 
Let us denote $x$ for the Galactic coordinate and $x'$ for the proper coordinate of detector. 
The density contrast is defined from the two point function of the scalar field $\langle \hat\phi^2(x) \rangle$, and since this quantity is a scalar under the general coordinate transformation, it should be the same in both frames, i.e. $\langle \hat\phi^2(x) \rangle = \langle \hat\phi'^2(x') \rangle$. 
Note that the density contrast is related to the total power of the signal. 
This indicates that the density contrast obtained in the previous section is directly related to the total power of the signal measured in detectors on Earth. 

The spectrum needs more careful discussion. 
The frequency $\omega$ is associated with the coordinate time $t$ of the Galactic frame. 
The Fourier transform in the Galactic frame is defined as $\hat \phi(\omega) = \int dt \, e^{i\omega t } \hat\phi(x)$, whereas it is defined as $\hat \phi' (\omega') = \int dt' \, e^{i\omega ' t'} \hat\phi(x)$ in the detector frame with a proper time $t'$. 
If the detector is an inertial observer moving in a constant velocity with respect to the Galactic frame, the proper coordinate can be related to the Galactic coordinate as $x' = \Lambda x$ with an appropriate Lorentz transformation $\Lambda$.
In this case, it is straightforward to show that the same expression for the spectrum, Eq.~\eqref{spec}, can be used in the detector frame upon a replacement of the velocity in the distribution with the velocity observed in the detector frame, $k \to \Lambda^{-1}k$. 

The orbital motion of the Earth introduces complications to the above discussion since a detector on Earth is not an inertial observer with a constant velocity. 
The above procedure of replacing $k \to \Lambda^{-1}k$ to obtain the spectrum in the detector frame is only valid if the time series is extended over a time scale sufficiently shorter than the orbital time scale of the Earth.
If one attempts to Fourier-transform the signal over a time scale longer than orbital time scale or the coherence time scale of wave dark matter, this procedure would introduce side-bands in the frequency space due to the frequency modulation induced by the orbital motion of Earth and the oscillations of the wave function.\footnote{For experiments sensitive to the gradient, the rotation of Earth introduces an amplitude modulation of the signal, leading to side-bands in the spectrum. This also introduces additional complications in the frequency space analysis~\cite{Lisanti:2021vij}.}

\subsection{Semiclassical limit}\label{sec:classical_limit}
In the previous section, we have observed that the wave result approaches the classical particle dark matter result when
\bea
m \tilde v (r) r \gg 1.
\label{cl_lim}
\eea
This is the limit at which the size of the system becomes much greater than the de Broglie wavelength of the wave dark matter. 
We provide a brief explanation why both results converge in this limit. 

The discussion of particle dark matter is entirely based on the phase space distribution modified by the gravitational potential of the source.
To discuss the limit at which particle and wave dark matter results converge, it is convenient to find a quantity analogous to the classical phase space distribution.
The quantum mechanical analog of the classical phase space distribution is given by the following quasi-probability distribution,
\bea
f_W(\bfx,\bfp) &=& \int d^3  y \, e^{i \bfp\cdot \bfy/\hbar}
\nonumber\\
& \times  &
\int d^3 v\, f(\bfv) \psi^*_\bfv (\bfx+\bfy/2) \psi_\bfv(\bfx-\bfy/2). 
\label{eq:wigner}
\eea
which was first introduced by Wigner~\cite{Wigner:1932eb}. 
We reintroduce the Planck constant $\hbar$. 
When it is integrated over the phase space $d^3 p$, it reproduces the variance of the field, which is directly related to the density contrast as
\bea
\int \frac{d^3p}{(2\pi)^3} f_W(\bfx,\bfp) = \int d^3v \, f(\bfv) |\psi_{\bfk}(\bfx)|^2 = \langle \phi^2\rangle. 
\eea
Moreover, the quasi-probability distribution $f_W$ evolves according to
\begin{eqnarray*}
\frac{\partial f_W}{\partial t} &+& \bfv \cdot \frac{\partial f_W}{\partial \bfx}  \\  
&+& \frac{i}{\hbar} \left[ V \big(\bfx + \frac{i\hbar}{2} \nabla_{\bfp} \big)- V\big(\bfx -\frac{i\hbar}{2} \nabla_{\bfp} \big) \right] f_W = 0. 
\end{eqnarray*}
This can be derived from the definition of $f_W$, Eq.~\eqref{eq:wigner}, and the \sch equation with potential $V$.
We see that the leading term in the semiclassical limit ($\hbar\to0$) is identical to the Boltzmann equation. 
From above, we can conclude that the wave and particle dark matter result should converge in the limit $\hbar \to0$. 
This semiclassical limit is identical to Eq.~\eqref{cl_lim}, which can be easily shown from the \sch equation. 
See Appendix~\ref{app:semiclassical} for more details.

\section{Conclusion}\label{sec:conclusion}
We have discussed the gravitational response of wave dark matter.
After a brief review on the gravitational response of particle dark matter, we have discussed and provided a simple formalism to investigate wave dark matter in the presence of the gravitational potential due to astrophysical object of mass $M$.
We specifically discuss a spinless bosonic field, minimally coupled to the gravity.
As in the canonical formulation of quantum field theories, we have expanded the dark matter field in terms of the creation and annihilation operators with a mode function. 
We solve the \sch equation to find the response of each wave mode and then specify the density operator which completely sets the statistical properties of the field. 
This procedure allows us to consider both the gravitational response and the stochastic nature of the wave dark matter.

A comparison of particle and wave dark matter reveals interesting similarity and at the same time differences between them. 
It is already clear from the monochromatic result that the wave nature becomes manifest only for $m\tilde v r \lesssim 1$, while for $m\tilde v r \gg1$, it practically converges to the result of particle dark matter. 
The characteristics of wave dark matter is the appearance of an additional scale, the de Broglie wavelength, $\tilde\lambda_{\rm dB}  = 2\pi / m\tilde v$. 
The spatial structure of the wave density contrast is smoothened on the scale of its de Broglie wavelength, which is the feature commonly observed in the studies of wave dark matter.

We have applied the formalism to the dark matter substructures as well as halo dark matter in the solar system. 
Different substructures are considered to examine the gravitational response of wave dark matter as a function of the mean velocity and the velocity dispersion.
Substructures considered here are motivated either by recent astrometric data or by a theoretical possibility to explain the stellar substructures in the Milky Way.
The wave characteristics are clearly seen in all of the provided examples: for $m\tilde v r \gg1$, the wave response is essentially the same as the particle response, while for $m \tilde v r \ll 1$, the wave result becomes independent on the location of the Earth. 
Interestingly, for $m\tilde v r\sim 1$, we observe that the wave density contrast could be at most a factor two larger than that of particle dark matter.
Moreover, for streams with small velocity dispersion, the angular distribution of the density contrast could still be different from that of particle density contrast. 
Although the gravitational focusing effects remain subleading in many examples we studied, they could provide important clues for the local dark matter structures upon the detection of dark matter.

\acknowledgements
This work is supported by the Deutsche Forschungsgemeinschaft under Germany’s Excellence Strategy - EXC 2121 Quantum Universe - 390833306.
We thank the hospitality of INFN Galileo Galilei Institute for Theoretical Physics during the workshop ‘New Physics from The Sky’.

%\newpage
\appendix

%% PARTICLE FOCUSING %%
\section{Particle gravitational focusing}\label{app:particle}
\begin{figure}
\includegraphics[width=0.45\textwidth]{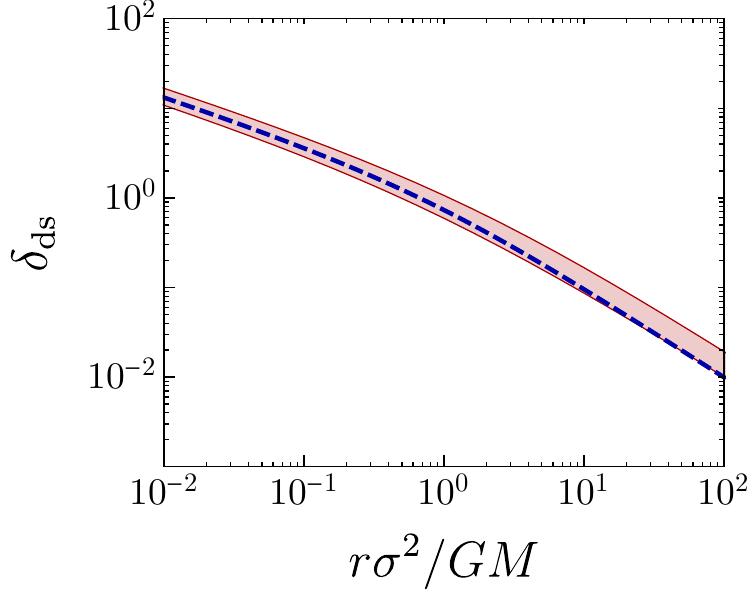}
\includegraphics[width=0.45\textwidth]{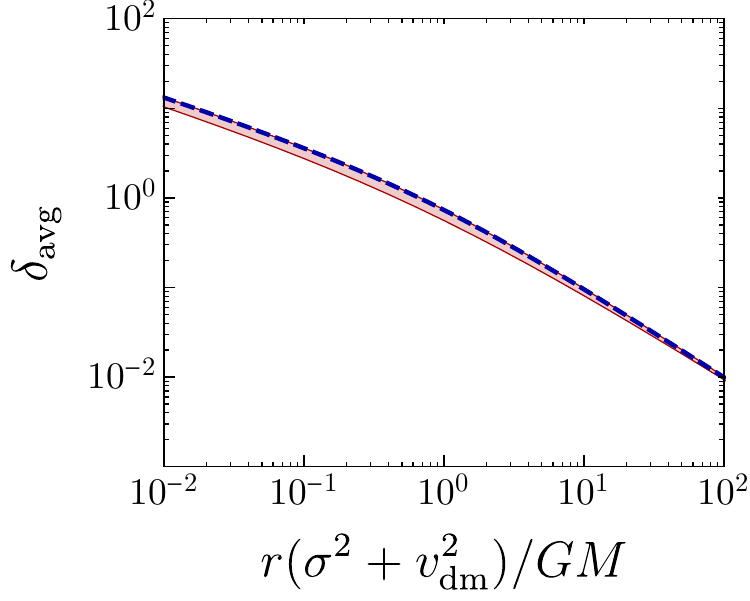}
\caption{%
(Top) We show the density contrast along the downstream $\mu = \hat{x}\cdot \hat{v}_{\rm dm} =1$ for different values of $v_{\rm dm} /\sigma \in [0.1, 10]$ (red band).
The lower and upper boundary correspond to $v_{\rm dm}/\sigma =0.1$, $10$, respectively.
The dashed line corresponds to Eq.~\eqref{particle_down_est}, $1+ \delta_{\rm ds} = \sqrt{1 + v_e^2/\sigma^2}$. 
The numerical result is consistent to the approximation Eq.~\eqref{particle_down_est} within a factor two for a wide range of $v_{\rm dm}/\sigma$. 
(Bottom) We show the density contrast averaged over the solid angle for $v_{\rm dm}/\sigma \in [0.1,10]$ (red band).
The lower and upper boundary correspond to $v_{\rm dm}/\sigma =0.1,\, 10$, respectively. 
The dashed line corresponds to the approximation Eq.~\eqref{particle_avg_est}.
The numerical result is consistent to the approximation Eq.~\eqref{particle_avg_est} within a factor two for a wide range of $v_{\rm dm}/\sigma$. }
\label{fig:particle_delta_properties}
\end{figure}

We discuss details of the gravitational response of particle dark matter.
In the main text, we have provided two approximations for (i) the density contrast along the downstream Eq.~\eqref{particle_down_est}
$$1 + \delta_{\rm ds} = 1+  \delta|_{\mu =1} \approx \Big( 1 + \frac{v_e^2}{\sigma^2} \Big)^{\frac12}$$ and (ii) the density contrast averaged over the solid angle Eq.~\eqref{particle_avg_est}
$$1 + \delta_{\rm avg} = \int \frac{d\Omega_{\hat x}}{4\pi} (1 + \delta) \approx \Big(1 + \frac{v_e^2}{\sigma^2 + v_{\rm dm}^2} \Big)^{\frac12}. $$
In Figure~\ref{fig:particle_delta_properties}, we compare these approximations and numerical results.
In the top panel, we plot $\delta_{\rm ds}$ for $v_{\rm dm}/ \sigma \in [ 0.1, 10]$. 
The result shows that $\delta_{\rm ds}$ does not crucially depend on the value of $v_{\rm dm}/\sigma$. 
The dashed line is the approximation given by Eq.~\eqref{particle_down_est}, and the numerical result matches well to the approximation within a factor of two for a wide range of $v_{\rm dm}/\sigma$.
In the bottom panel, we plot $\delta_{\rm avg}$ for the same range of $v_{\rm dm}/\sigma$. 
The dashed line is the approximation given by Eq.~\eqref{particle_avg_est}.
The result $\delta_{\rm avg}$ is only sensitive to $r (\sigma^2 + v_{\rm dm}^2)/GM$.

In the limit $v_e \ll \sigma$, an analytic expression for the density contrast can be obtained~\cite{2008gady.book.....B},
$$
\delta(\bfx) \approx \frac{v_e^2}{2\sigma^2} 
\exp\left[ - \frac{v_{\rm dm}^2}{2\sigma^2} \sin^2\theta \right]
{\rm erfc}\left( - \frac{v_{\rm dm}}{\sqrt{2}\sigma} \cos\theta \right)
$$
where $\cos\theta = \hat{x}\cdot \hat{v}_{\rm dm}$. 
From this, one may obtain the approximations Eqs.~\eqref{particle_down_est}--\eqref{particle_avg_est} for $v_e / \sigma <1$. 
This expression can be used for an investigation of dark matter substructures with $\sigma > v_e$.

%APPENDIX: WAVE FOCUSING %
\section{Wave gravitational focusing}\label{app:wave}

\subsection{Average density contrast in monochromatic limit}
\begin{figure}[t]
\centering
\includegraphics[width=0.45\textwidth]{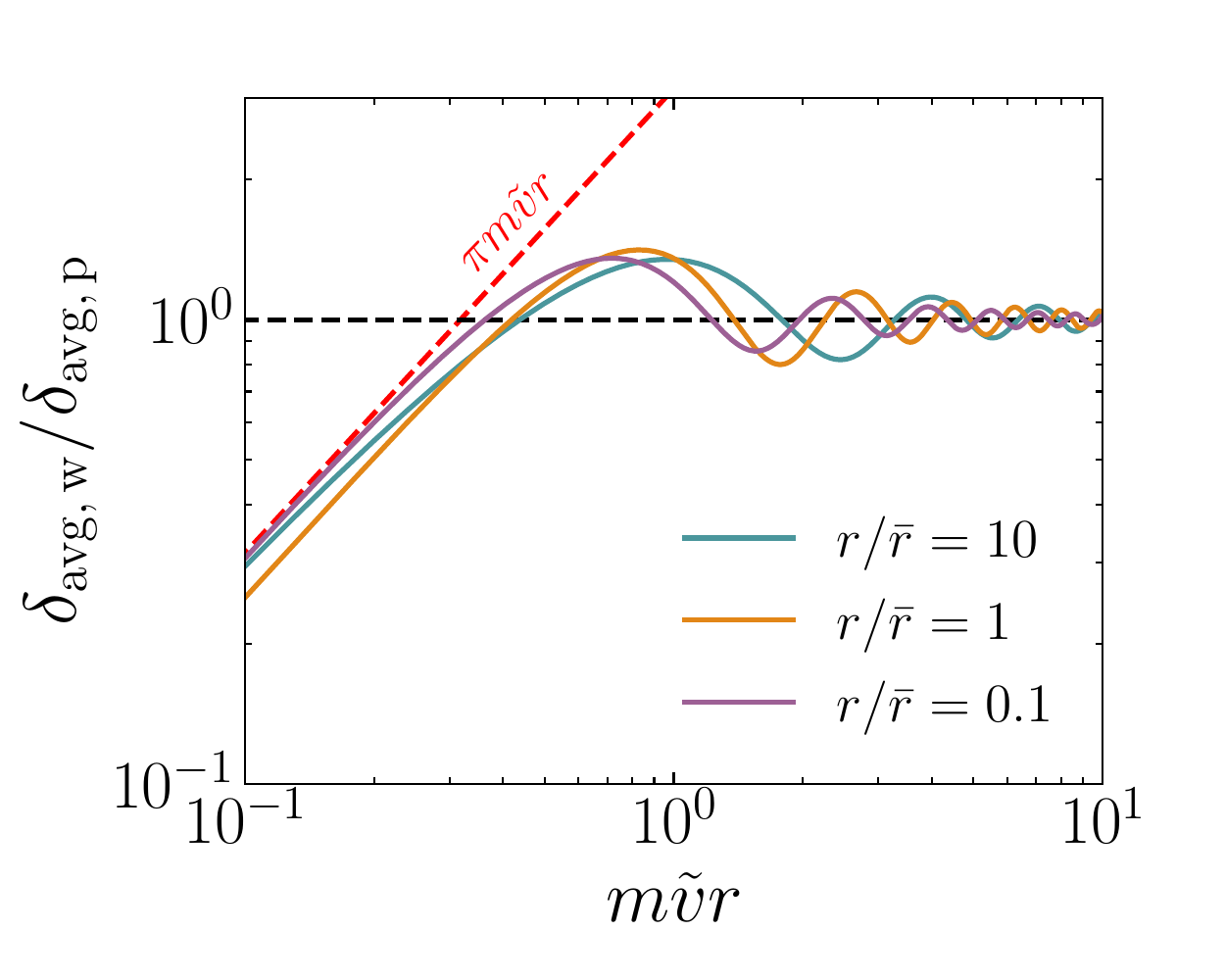}
\caption{%
We present the density contrast averaged over the solid angle in the monochromatic case.
Different solid lines correspond to different values of $r / \bar{r}$. 
All of them approach to the classical particle result for $m \tilde v r \gg 1$.
For $m \tilde v r \lesssim 1$, the averaged density contrast is suppressed. }
\label{fig:wave_mono_3}
\end{figure}

In Figure~\ref{fig:wave_mono_3}, we present the density contrast averaged over the solid angle $\hat{x}$ in the monochromatic limit for different values of $r/\bar r$. 
The averaged density contrast is normalized with $ \delta_{\rm avg, p}(r) = \sqrt{1 + v_e^2(r) / v^2} +1$. 
We observe that for $m\tilde v r$ the wave dark matter result is suppressed relative to the particle dark matter result as $\delta_{\rm avg, w} / \delta_{\rm  avg,p} \propto  \pi m \tilde v r$. 
The behavior of $\delta_{\rm  avg, w} / \delta_{\rm avg, p}$ does not sensitively depend on the values of $r / \bar{r}$.

\subsection{Semiclassical limit}\label{app:semiclassical}
In this appendix, we discuss the semiclassical limit in the \sch equation. 
We reintroduce the Planck constant $\hbar$. 
We begin with the \sch equation, $i \hbar \dot{\psi} = (-\hbar^2 \nabla^2 /2m + V)\psi$. 
With $\psi(t,\bfx) \to e^{-i E t} \psi (\bfx)$, the \sch equation becomes
\bea
0 = \left[ \nabla^2 + \frac{2m(E - V)}{\hbar^2}  \right] \psi.
\eea
The energy eigenvalue is $E = k^2/2m$ where $\bfk$ is the wave number asymptotically far way from the source $M$. 
The wave function is characterized by $\bfk$. 
We perform Legendre expansion of the field, 
$$\psi_{\bfk} (\bfx) = \frac{1}{r} \sum_{\ell = 0}^\infty \chi_{k\ell}(r) P_\ell (\cos\theta),$$ where $\cos\theta =\hat{x}\cdot\hat{v}$. 
This procedure leads to the \sch equation for the radial wave function $\chi_{k\ell}$,
\bea
\chi_{k\ell}''+ \frac{2m(E- V_\ell)}{\hbar^2} 
 \chi_{k\ell} = 0,
\eea
where the prime denotes the derivative with respect to $r$ and the potential $V_\ell$ is defined as
\bea
V_\ell (r) = V(r) + \frac{\hbar^2}{2m }  \frac{ \ell(\ell+1)}{r^2} 
= V(r) + \frac{L^2}{2m r^2} .
\eea
We have defined $L^2 = \hbar^2 \ell (\ell +1)$.

We seek for the solution of the following form, 
\bea
\chi_{k\ell} (r) = e^{i \sigma / \hbar} = \exp\left[ \frac{i}{\hbar} \sum \hbar^n \sigma^{(n)}_\ell \right]. 
\eea
The \sch becomes $\sigma'^2 - i \hbar \sigma'' = 2m (E - V_\ell)$. 
By solving the \sch equation order by order in $\hbar$, we find 
\bea
\sigma_\ell^{(0)}(r) &=&  \int^r dr'\, \sqrt{2m [E - V_\ell]}  \equiv \int^r dr' \, p_\ell(r') ,
\\
\sigma_\ell^{(1)}(r) &=& \frac{i}{2} \ln p_\ell(r).
\eea
We have defined $p_\ell(r) \equiv \sqrt{2m [ E - V_\ell (r)] }$, which is identical to the radial momentum in $1/r$-potential in classical mechanics. 

This semi-classical or WKB approximation is valid when $|\hbar p_\ell' / p_\ell^2 | \ll 1$. 
This can be written as
\bea
\frac{d}{dr} \frac{\hbar}{p_\ell}  \ll 1.
\eea
Naively speaking, this condition is approximately $\hbar/ p_{\ell} r \sim \lambda_{\rm dB} / r \ll1$, indicating that the semi-classical limit is valid only when the size of the system is much greater than the de Broglie wavelength, $\lambda_{\rm dB} \sim 1/p$. 
More specifically, the above condition can be written as
\bea
\frac{\hbar m |F_\ell|}{p_\ell^3} \ll1
\label{wkb_validity}
\eea
with
\bea
F_{\ell} =p_\ell' = \frac{1}{2m r^3} \left[ (m v_e r)^2 - 2 L^2 \right]
\eea
For a given set of parameters, the dominant contribution in the series expansion of wave function arises at $\ell \sim m \tilde v r$.
With $m \tilde v r > m v_e r$, it is straightforward to show that this condition Eq.~\eqref{wkb_validity} can be approximated as $m\tilde v r \gg \hbar$. 

\begin{figure}[t]
\vspace{1cm}
\centering
\includegraphics[width=0.48\textwidth]{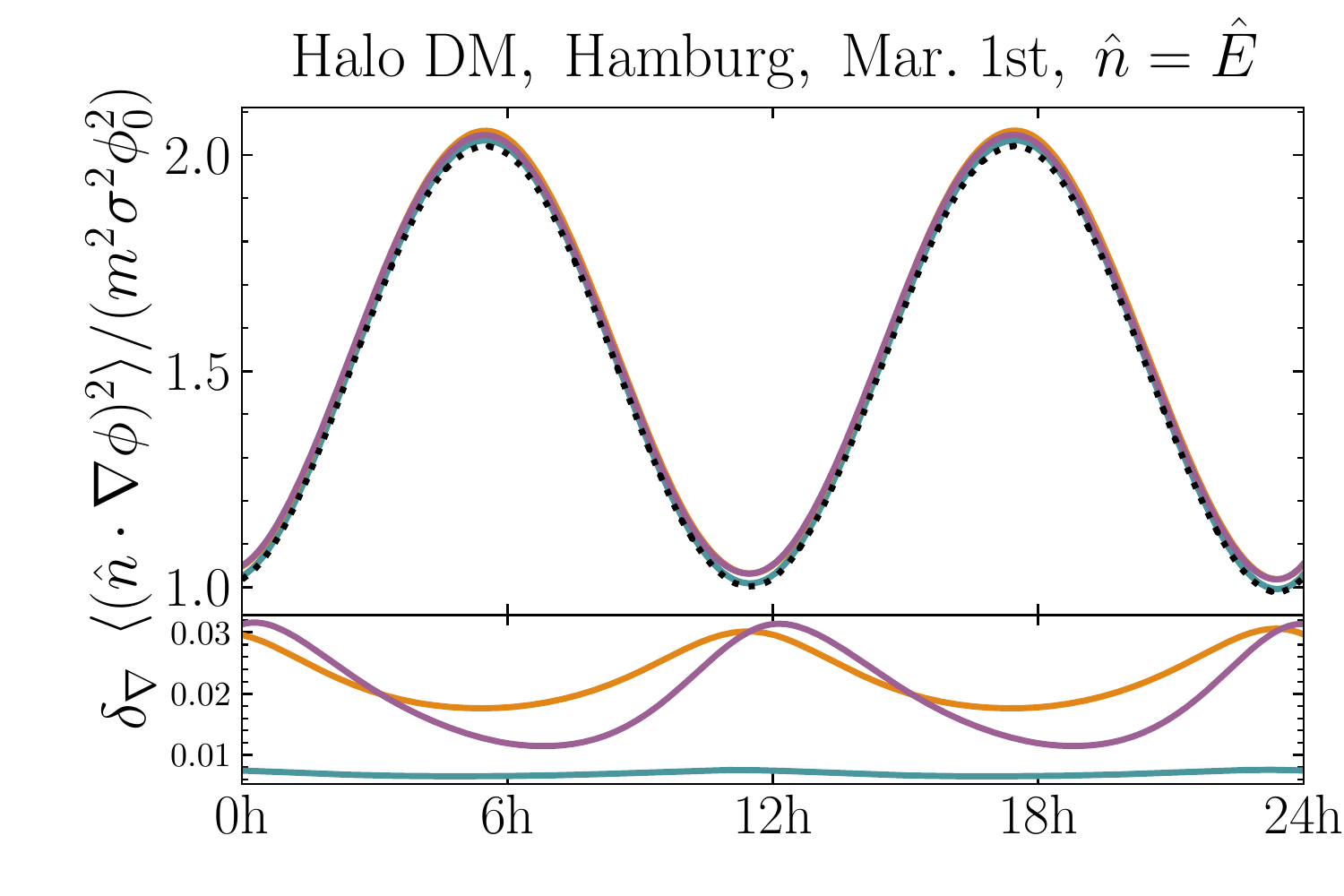}
\includegraphics[width=0.48\textwidth]{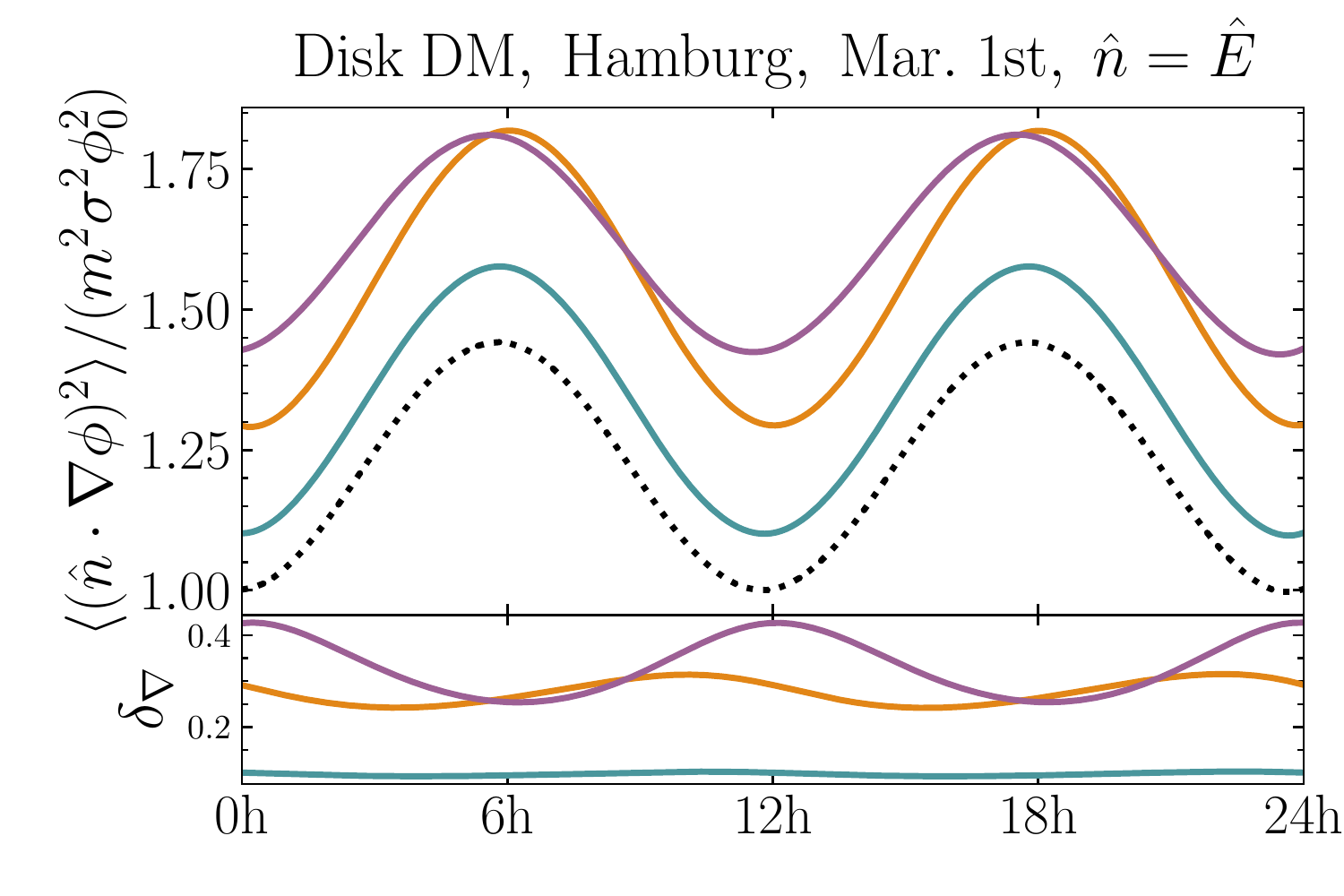}
\caption{Daily modulation of the variance of the field gradient projected on the sensitivity axis of a detector located in Hamburg, Germany and pointing in the east direction for halo DM (top) and dark disk (bottom) on March 1st, 2021. Colored lines are the results with gravitational focusing focusing for different values of $\tilde{\lambda}_{\rm dB}$ as in Figure \ref{fig:halo} (top) and \ref{fig:disk} (bottom), respectively. The black dotted line denotes the asymptotic result. The bottom panels shows the contrast of focused result with respect to unfocused ones. }
\label{fig:gradient}
\end{figure}

\section{Field gradient daily modulation}\label{app:gradient}
In Figure~\ref{fig:gradient}, we show the daily modulation of the field gradient calculated with Eq.~\eqref{gradphi}. As a benchmark we consider a detector located in Hamburg, Germany $(54^\circ {\rm N},10^\circ {\rm E} )$ with sensitivity axis pointing towards the east direction on March 1st, 2021. We show the result for halo DM and dark disk component, with same benchmark choices for $m\tilde{v}r$ exploited in Figure \ref{fig:halo} (top) and \ref{fig:disk} (bottom). The dotted black line shows the result with no focusing, asymptotically far away from the solar system. In the bottom panels we show the contrast $\delta_\nabla$ between the focused lines and the asymptotic value. Note that in both cases the magnitude of the contrast is comparable to the density contrast, signaling the effect is of order $\mathcal{O}(v_e^2/\sigma^2)$. Another interesting effect due to gravitational focusing is the modification of the position of the peaks. This effect is due to the velocity component acquired by DM from the gravitation potential of the Sun, which changes the direction of the total DM velocity vector, hence the time at which the peak value is reached. 

\newpage 

\bibliography{ref}
\end{document}